\documentclass[10pt]{article}
\pdfoutput=1
\usepackage[pdftex]{color}
\usepackage{amssymb}
\usepackage{amsthm}
\usepackage{amsmath}
\usepackage{latexsym}
\usepackage{amscd}
\usepackage{graphicx}
\usepackage{setspace}
\usepackage{cancel}
\usepackage{enumerate}

\renewcommand{\proof}[1]{\noindent \normalfont{\textbf{Proof.} \  \  #1 \hfill $\Box$}\\}

\newtheorem{thm}{Theorem}[section]
\newtheorem{prop}[thm]{Proposition}
\newtheorem{lemma}[thm]{Lemma}
\newtheorem{corol}[thm]{Corollary}

\newcommand{\ds}{\displaystyle}

\begin{document}

\title{Hebbian Crosstalk and Input Segregation}

\author{Anca R\v{a}dulescu$^{1}$, Paul Adams$^{2,3}$}
\maketitle

\noindent $^1$Department of Mathematics, 395 UCB, University of Colorado, Boulder, \emph{radulesc@colorado.edu}\\

\noindent $^2$Department of Neurobiology and Behavior, Stony Brook University, Stony Brook, \emph{padams@notes.cc.sunysb.edu}\\

\noindent $^3$Kalypso Institute, Stony Brook, NY 11790

\begin{abstract}

\noindent {\bf Purpose.} We previously proposed that Hebbian adjustments that are incompletely synapse specific (``crosstalk'') might be analogous to genetic mutations. We analyze aspects of the effect of crosstalk in Hebbian learning using the classical Oja model.\\

\noindent {\bf Methods.} In previous work we showed that crosstalk leads to learning of the principal eigenvector of $\bf{EC}$ (the input covariance matrix pre-multiplied by an error matrix that describes the crosstalk pattern), and found that with positive input correlations increasing crosstalk smoothly degrades performance. However, the Oja model requires negative input correlations to account for biological ocular segregation. Although this assumption is biologically somewhat implausible, it captures features that are seen in more complex models. Here, we analyze how crosstalk would affect such segregation.\\

\noindent {\bf Results.} We show that for statistically unbiased inputs crosstalk induces a bifurcation from segregating to non-segregating outcomes at a critical value which depends on correlations. We also investigate the behavior in the vicinity of this critical state and for weakly biased inputs.\\

\noindent {\bf Conclusions.} Our results show that crosstalk can induce a bifurcation under special conditions even in the simplest Hebbian models and that even the low levels of crosstalk observed in the brain could prevent normal development. However, during learning pairwise input statistics are more complex and crosstalk-induced bifurcations may not occur in the Oja model. Such bifurcations would be analogous to ``error catastrophes'' in genetic models, and we argue that they are usually absent for simple linear Hebbian learning because such learning is only driven by pairwise correlations.

\end{abstract}

\vspace{1cm}



\clearpage
\section{Introduction}

\subsection{Background}

Learning is thought to occur as a result of changes in synaptic strength triggered by pre- and postsynaptic neural activity, in a ``Hebbian'' manner. Such changes are not completely specific to the synapses at which the activity occurs~\cite{harvey2007locally,bi2002spatiotemporal,bonhoeffer1989synaptic}, because of inevitable albeit minimal second-messenger diffusion.

Oja~\cite{oja1982simplified} showed that a simple model neuron could perform  unsupervised Hebbian learning of the first principal component of an input distribution. In this model, unlimited weight growth is prevented using an additional term in the learning rule, producing an implicit, ``multiplicative'' weight normalization ~\cite{malsburg1973self}. Biological synapses do show Hebbian properties, using well-understood, spike-coincidence detection machinery, raising the possibility that real neurons can exhibit similar unsupervised learning. Finding principal components could be very useful in the brain for data compression and transmission, since for Gaussian data such representations have statistically optimal properties, and often neural signals are approximately Gaussian. Furthermore, representational learning often requires that inputs be pairwise decorrelated. Hebbian learning can also explain developmental changes, such as the segregation of visual input to central neurons.

Recent data suggest~\cite{harvey2007locally,bi2002spatiotemporal,bonhoeffer1989synaptic} that weight updates may be affected by each other, for example due to unavoidable residual second messenger diffusion between closely spaced synapses. We have suggested that such crosstalk is analogous to mutation in genetics, and that cortical circuitry may be specialized to reduce it. However, it is not clear that learning would be subject to an ``error catastrophe'' such as that occurring in genetic systems~\cite{eigen1971selforganization}. If complete learning failure does not occur at a critical, low, crosstalk level, such circuitry might not be necessary.

In a recent paper~\cite{radulescu2009hebbian} we examined how crosstalk would affect the Oja model. We considered a learning network consisting of a single output neuron receiving, through a set of $n$ input neurons, $n$ signals ${\bf x} = (x_1,...,x_n)^T$ drawn from a probability distribution ${\cal{P}}({\bf x}), \; {\bf x} \in \mathbb{R}^{n}$, transmitted via synaptic connections of strengths $\mbox{\boldmath $\omega$} = (\omega_1,...,\omega_n)^T$. The resulting scalar output $y$ was generated as the weighted sum of the inputs $y={\bf x}^T\mbox{\boldmath $\omega$}$.

The synaptic weights $\omega_i$ were modified in accordance with Oja's rule of learning, by implementing first a Hebb-like strengthening proportionally with the product of $x_i$ and $y$ (with small constant of proportionality, or \emph{learning rate}, $\gamma$)

 $$\omega_{i}(t+1) = \omega_{i}(t) + \gamma y(t) x_{i}(t)$$

\noindent followed by an approximate ``normalization'' step (applicable for small $\gamma$ and $\| {\bf w} \|$ close to one), maintaining the Euclidean norm of the weight vector $\mbox{\boldmath $\omega$} = (\omega_1,...,\omega_n)^T$ close to one.

$$\mbox{\boldmath $\omega$}(t+1) = \mbox{\boldmath $\omega$}(t)+\gamma y(t)[{\bf x}(t)- y(t)\mbox{\boldmath $\omega$}(t)]$$

We considered the long-term average of this Oja equation, using the input covariance matrix ${\bf C}={\bf x}^T{\bf x}$ as an appropriate long-term characterization of the inputs, and studying the behavior of ${\bf w}(t)= \langle \mbox{\boldmath $\omega$}(t+1) \vert \mbox{\boldmath $\omega$}(t) \rangle$:

$$ \frac{d{\bf w}}{dt} = \gamma \left[ {\bf C w} - \left( {\bf w}^{T}{\bf C w} \right) {\bf w} \right]$$

\noindent in continuous time, or:

$$ \Delta{\bf w} = \gamma \left[ {\bf C w} - \left( {\bf w}^{T}{\bf C w} \right) {\bf w} \right]$$

\noindent as a discrete time approximation.

We then introduced inspecificity into the learning equation~\cite{radulescu2009hebbian}. We implemented this inspecificity by assuming that, on average, only a fraction $q$ of the intended update reaches the appropriate connection, the remaining fraction $1-q$ being distributed amongst the other connections (according to a rule which we defined according to plausible underlying biology). The quality factor $q$ is analogous to a similar factor in molecular evolution theory that represents the fidelity of single-base copying~\cite{swetina1982self}. The actual update at a given connection thus includes contributions from erroneous or inaccurate updates from other connections. The erroneous updating process was formally described by an error matrix  $\mbox{\boldmath ${\cal E}$}$, independent of the inputs, whose elements, which depend on average on $q$, reflect at each time step $t$ the fractional contribution that the activity through the connection with weight $\omega_{i}$ makes to the update of $\omega_{j}$.

$$\omega_{i}(t+1) = \omega_{i} + \gamma y([\mbox{\boldmath ${\cal E}$}{\bf x}]_{i}- y \omega_{i})$$

\noindent The discrete long-term statistics can be then written in matrix form as:

$$\Delta {\bf w} = \gamma \left[ {\bf ECw}- \left( {\bf w}^{T}{\bf Cw} \right) {\bf w} \right]$$

\noindent where the ``error matrix'' ${\bf E}= \langle \mbox{\boldmath ${\cal E}$} \rangle$ is a symmetric matrix with positive entries, which equals the identity matrix ${\bf }I \in {\cal{M}}_{n}(\mathbb{R})$ in case of perfect quality updates. Then the rule changes into:

\begin{equation}
\frac{d{\bf w}}{dt} = \gamma \left[ {\bf EC w} - \left( {\bf w}^{T}{\bf C w} \right) {\bf w} \right]
\label{mothersys}
\end{equation}

Throughout the paper, we call this the (inspecific) Oja rule with continuous time updates.


\noindent We studied the asymptotic behavior of this n-dimensional system, starting with a local linear analysis of the equilibria and their stability. Although this rule is nonlinear, the Hebbian update term is linear in the output, and we sometimes refer to this, and related, rules, as being ``linear,'' in contrast to other Hebbian rules~\cite{hyvarinen1998independent,hyvarinen2001independent,bell1995information,olshausen1996emergence,elliott2003analysis,foldiak1990forming,cooper2004theory} which are nonlinear in the output.

Note that the symmetric, positive definite matrix ${\bf C} \in {\cal{M}}_{n}(\mathbb{R})$ defines a dot product between any two vectors ${\bf w}$ and ${\bf v}$ in $\mathbb{R}^{n}$ as $\ds \langle {\bf v},{\bf w} \rangle_{\bf C}={\bf v}^{T}{\bf Cw}$. Although both ${\bf C}$ and ${\bf E}$ are symmetric, the product ${\bf EC}$ is not symmetric in the Euclidean metric. However, in a new metric defined by the dot product $\langle \cdot,\cdot \rangle_{\bf C}$, ${\bf EC}$ is symmetric: $\ds \langle {\bf ECu},{\bf v} \rangle_{\bf C} = ({\bf ECu})^t{\bf Cv} = {\bf u}^t {\bf C}^t {\bf E}^t {\bf Cv} = {\bf u}^t {\bf CECv} = \langle {\bf u}, {\bf ECv}\rangle_{\bf C} \text{ , for all } {\bf u},{\bf v} \in \mathbb{R}^n$. Hence ${\bf EC}$ has a basis of eigenvectors, orthogonal with respect to the dot product $\langle \cdot,\cdot \rangle_{\bf C}$. The following is immediate:

\vspace{3mm}
\noindent {\bf Description of equilibria of the system (\ref{mothersys}).} \emph{An equilibrium for the system is any vector ${\bf w}=(w_{1}...w_{n})^{T}$ such that ${\bf ECw}=({\bf w}^{T}{\bf Cw}){\bf w}$, i.e., an eigenvector of ${\bf EC}$ (with corresponding eigenvalue $\lambda_{\bf w}$), normalized, with respect to the norm $\lVert \cdot \rVert_{\bf C} = \langle \cdot,\cdot \rangle_{\bf C}$, so that $\lVert {\bf w} \rVert_{\bf C}=\lambda_{\bf w}$. Generically, ${\bf EC}$ has a strictly positive, unique maximal eigenvalue, and the corresponding eigendirection is orthogonal in $\langle \langle \cdot,\cdot \rangle \rangle_{\bf C}$ to all other eigenvectors of ${\bf EC}$.}

\vspace{3mm}
\noindent For an equilibrium ${\bf w}$ of the system (\ref{mothersys}), the Jacobian matrix $Df^{\bf E}_{\bf w}$ around ${\bf w}$ is (see Appendix 1):

\begin{equation}
Df^{\bf E}_{\bf w} = \gamma \left[ {\bf EC} - 2{\bf w}({\bf Cw})^{T} - ({\bf w}^{T}{\bf Cw}){\bf I} \right]
\end{equation}

Then we have the following (see Appendix 1 for proof):

\vspace{3mm}
\noindent {\bf Stability criteria for equilibria.} \emph{Suppose ${\bf EC}$ has a multiplicity one largest eigenvalue. A normalized eigenvector ${\bf w}$ is a local hyperbolic attracting equilibrium for (\ref{mothersys}) iff it corresponds to the maximal eigenvalue of ${\bf EC}$.}

\vspace{3mm}
\noindent Such attractors always exist provided ${\bf EC}$ has a maximal eigenvalue of multiplicity one, which is generically true. Then the network learns, depending on its initial state, one of the two stable equilibria, which are the two (opposite) maximal eigenvectors of the modified input distribution, normalized so that $\| {\bf w} \|_{\bf C} = \lambda_{\bf w}$. It can be shown easily that these two attractors (the appropriately normalized eigenvectors corresponding to the maximal eigenvalue of ${\bf EC}$) can be the only attractors in the system (see Appendix 2 for proof).\\


\noindent In a previous paper~\cite{radulescu2009hebbian}, we further analyzed the sensitivity of the system under variations of parameters, for some biologically plausible forms of the covariance and error matrices:

\vspace{3mm}
${\bf C}=\left[\begin{array}{cccc}
v+\delta_1&c&\cdots &c\\
c&v+\delta_2&\cdots &c\\
\vdots & &\ddots &\vdots \\
c&c&\cdots &v+\delta_n\\
\end{array}\right]$ and
${\bf E}=\left[\begin{array}{cccc}
q&\epsilon&\cdots &\epsilon\\
\epsilon&q&\cdots &\epsilon\\
\vdots & &\ddots &\vdots \\
\epsilon&\epsilon&\cdots &q\\
\end{array}\right]$\\

\vspace{3mm}
\noindent where the input covariance matrix had uniform covariances $c>0$ and variance biases $\delta_1 \geq \delta_2 \geq \hdots \geq \delta_n$; the error matrix was defined such that $q> \epsilon > 0$, $q+(n-1)\epsilon=1$. Our analysis of this system concluded that the effect of biologically realistic levels of crosstalk would typically only produce small gradual changes in the learning process, though when inputs carry very similar signals, the effects could be more dramatic. In this paper we explore this ``very similar'' scenario more thoroughly. In particular we describe the effect of crosstalk in the special ``unbiased'' case, where the inputs have identical statistics.

\subsection{Biased and unbiased inputs}

Our previous analysis considered only distributions of inputs with a bias in the covariance matrix (we imposed the condition that ${\bf EC}$ has a leading eigenvalue of multiplicity one). While this case is mathematically generic, previous work using related models (without crosstalk~\cite{miller1989ocular}) to study learning in the visual system, often assumed that the input statistics are ``unbiased,'' or identical for each input (for example, because inputs from corresponding points in the left and right eyes see the same point in visual space). It is well known that in the two-dimensional case, if the two inputs $x_1$ and $x_2$ are positively correlated (as one might anticipate for active vision), linear Hebbian learning does not predict the observed developmental segregation of visual afferents~\cite{dayan2001theoretical,cooper2004theory,swindale1996development,willshaw1976patterned}; negative correlations (or a nonlinear rule) are required. However, modifications in learning rules, for example subtractive normalization ~\cite{goodhill1993topography,willshaw1976patterned,miller1994role,linsker1986basic}, a weight-dependent rule~\cite{elliott2002multiplicative} or a BCM rule~\cite{cooper2004theory}, although not always originally developed to explain segregation, can overcome this difficulty. Often these rules lead to Hebbian learning driven by a modified version of the covariance matrix. In the current work, we examine the dynamics of Oja learning with crosstalk when inputs are unbiased, and how this changes when a slight bias is introduced.

We show here that in the unbiased negative correlation case, the system undergoes a bifurcation in dynamics at a critical crosstalk level. Related results have been obtained by~\cite{elliott2012,cox2009hebbian}. While there is no true bifurcation in the near-unbiased case, the very dramatic change in learning that occurs over a small error range is biologically indistinguishable from a true bifurcation. We discuss our results in relation to models of development and learning.

\subsection{Our current model}

In this paper, we will consider the continuous-time, two-dimensional nonlinear rule of Oja (i.e., for two input channels and one output), with covariance matrix ${\bf C}$ and error matrix ${\bf E}$ symmetric matrices having the forms
$\displaystyle{{\bf C}=\left( \begin{array}{cc}
       v+\delta & c \\
       c & v
         \end{array} \right)}$ and
$\displaystyle{{\bf E}=\left( \begin{array}{cc}
       q & 1-q \\
       1-q & q
         \end{array} \right)}$.
The parameters are such that $1/2 < q \leq 1$, $v > 0$ and $c<0$, such that $v>\lvert c \rvert$, $v>\lvert \delta \rvert$ and $v(v+\delta)>c^2$ (i.e., $\det({\bf C})>0$). The 2D system expands to:

\begin{eqnarray}
\dot{w_1} &=& [q(v+\delta)+(1-q)c]w_1+[qc+(1-q)v]w_2 - [vw_1^2+2cw_1w_2+vw_2^2]w_1 \nonumber \\
\dot{w_2} &=& [(1-q)(v+\delta)+qc]w_1+[(1-q)c+qv]w_2 - [vw_1^2+2cw_1w_2+vw_2^2]w_2
\label{example}
\end{eqnarray}

\vspace{3mm}
The rest of the paper is centered around this 2-dimensional model. In Section 2, we establish the mathematical background of the model's behavior. We analyze some of its local and global dynamics, observe the dependence of these dynamics on parameters and discuss bifurcations. One of the phenomena central to our interest is how the behavior of the system changes when the bias parameter $\delta$ varies, in particular when it approaches zero (i.e., the inputs are very close to a perfectly unbiased state). In Section 3, we discuss the results in the context of visual modeling and ocular segregation of inputs.

\section{Linear analysis of the 2D dynamics}

We notice that the phase plane of the system is symmetric about the origin (i.e., if $w(t)$ is a solution curve for the system, then $-w(t)$ is as well). The trace, determinant and eigenvalues of ${\bf EC}$ can be obtained easily as expressions of the system parameters:

$\quad \det({\bf EC})=\det({\bf E}) \det({\bf C})=(2q-1)[v(v+\delta)-c^2]>0$\\

$\quad \text{tr}({\bf EC})=2(1-q)c+q(2v+\delta)>0$ (from the Cauchy-Schwartz inequality).

\begin{lemma}
For all parameter values, ${\bf EC}$ has two real eigenvalues $\mu_{1,2}$, which are distinct unless the conditions $\delta=0$ and $\displaystyle{q=q^*=\frac{v}{v-c}}$ are simultaneously satisfied. More precisely, when $\mu_1 \neq \mu_2$, we have:\\

\noindent $\displaystyle{\mu_1=\frac{2(1-q)c+q(2v+\delta)+\sqrt{\Delta}}{2}}$ larger eigenvalue, with eigenline of slope $\displaystyle{z_1=\frac{-q\delta+\sqrt{\Delta}}{2 \beta}}$\\

\noindent $\displaystyle{\mu_2=\frac{2(1-q)c+q(2v+\delta)-\sqrt{\Delta}}{2}}$ smaller eigenvalue, with eigenline of slope $\displaystyle{z_2=\frac{-q\delta-\sqrt{\Delta}}{2 \beta}}$\\

\noindent where $\beta=qc+(1-q)v$ and $\Delta=[2qc+(1-q)(2v+\delta)]^2+(2q-1)\delta^2$.
\end{lemma}

\begin{figure}[h!]
\includegraphics[scale=0.16]{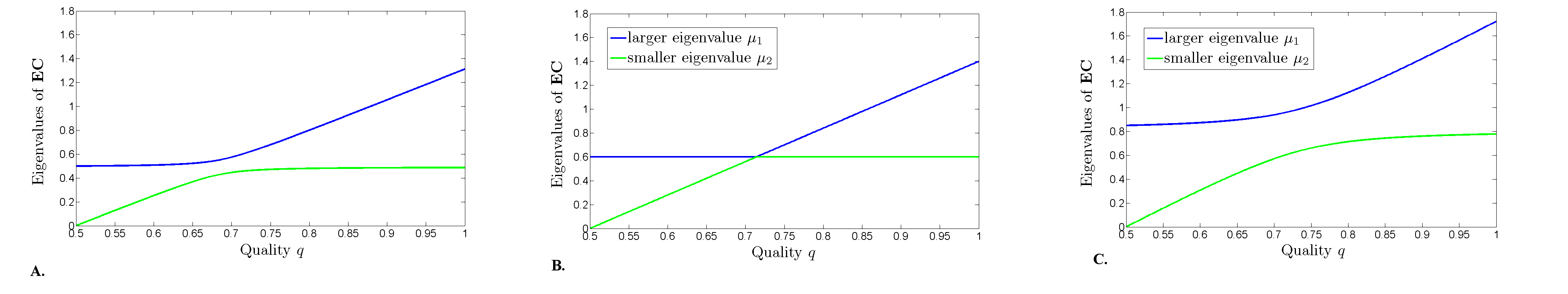}
\caption{{\bf Evolution of the eigenvalues as the quality $q$ is varied, in three different $\delta$ slices.} $\delta=-0.2$ ({\bf A}), $\delta=0$ ({\bf B}) and $\delta=0.5$ ({\bf C}). Fixed parameters: $v=1$ and $c=-0.4$, hence $q^*=1/1.4 ~\sim 0.71$. When $\delta=0$, the eigenvalues $\mu_1$ and $\mu_2$ touch at $q=q^*$. For $\delta \neq 0$, the two curves avoid this crossing; the minimal distance between them occurs at $q=q^*$, but it is strictly positive.}
\label{qvaried}
\end{figure}

\proof{ The calculation of eigenvalues and eigenvectors is immediate from the characteristic equation of ${\bf EC}$: $X^2-\text{tr}({\bf EC})X+\det({\bf EC})=0$, with discriminant

$$\Delta=\text{tr}({\bf EC})^2-4\det({\bf EC})=[2qc+(1-q)(2v+\delta)]^2+(2q-1)\delta^2$$

\noindent Notice that $\Delta \geq 0$, with equality $\Delta=0$ (i.e., double eigenvalue for ${\bf EC}$) iff both $\delta=0$ and $\beta=0$. The critical quality value (where the two eigenvalues are equal, producing a switch in the dynamics when $\delta=0$) is $\displaystyle{q^*=\frac{v}{v-c}}$. Since $v > \lvert c \rvert$, this value occurs within the appropriate $q$ range, $(1/2,1]$ (see Figure~\ref{qvaried}).}

\subsection{Equilibria of the 2D system}

Throughout this section, in addition to working in our generally specified parameter ranges, we will assume that ${\bf EC}$ has distinct eigenvalues (i.e., $\delta \neq 0$ or $q \neq q^*$). In this case, the system has as equilibria the origin ${\bf w}=0$, and two pairs of opposite eigenvectors of ${\bf EC}$ normalized such that ${\bf w}^t{\bf Cw}=\mu$ (where $\mu$ is the respective eigenvalues of each pair).\\

The normalization condition can be written as:

$${\bf w}^t{\bf Cw} = (v+\delta)w_1^2+2cw_1w_2+vw_2^2=\mu$$

\noindent Using the same notation $z=w_2/w_1$, this can be rewritten as $vz^2+2cz+(v+\delta)=\mu/w_1^2$, so that

$$\| {\bf w} \| = \sqrt{\frac{\mu(z^2+1)}{vz^2+2cz+(v+\delta)}}$$

\noindent Thus the norm varies with both error and correlation. The position and stability of the four nonzero equilibria vary with the parameters $v,c,\delta$ and $q$. If we aim to study the sensitivity of the system's dynamics under parameter perturbations, the next step should be establishing the linear stability of these equilibria; this follows directly from the general results in Section 1:

\vspace{3mm}
\noindent {\bf Description and stability of equilibria.} \emph{Suppose the matrix ${\bf EC}$ has distinct eigenvalues. The system (~\ref{example}) has five distinct equilibria, ${\bf w}=0$ and four normalized eigenvectors of ${\bf EC}$. The two (opposite) eigenvectors of the larger eigenvalue are hyperbolic attractors, and the two (opposite) eigenvectors corresponding to the lower eigenvalue are saddles. The origin is repelling.}

\vspace{3mm}
\noindent More precisely, this means that if $\mu_{\bf w}$ is the larger eigenvalue of ${\bf EC}$, the Jacobian matrix $D_{\bf w}$ has two negative eigenvalues, hence ${\bf w}$ is an attracting node. If instead $\mu_{\bf w}$ is the smaller eigenvalue of ${\bf EC}$, then $D_{\bf w}$ has two real eigenvalues of opposite signs, and ${\bf w}$ is a saddle equilibrium.\\

\vspace{3mm}
\noindent We are particularly interested in the behavior near and at $\delta=0$. The above characterization of equilibria applies when $\delta \neq 0$, but it breaks down in the parameter slice $\delta=0$, at the critical point when ${\bf EC}$ has a double eigenvalue. In other words we expect that the system undergoes a bifurcation in the unbiased $\delta=0$ slice, which does not exist in the other, $\delta \neq 0$ slices (i.e., when ``bias'' is present in the inputs), therefore we will study this case separately.\\

\noindent For the following paragraph (Section 2.2) we assume $\delta \neq 0$. The unbiased case $\delta=0$ is discussed separately in Section 2.3. The results are integrated and concluded in Section 2.4.

\subsection{More properties of the phase plane}

One way to describe the dynamics of the system, including the more global aspects and possibly cyclic behavior (which has not yet been excluded) is to follow the rotational direction of the solution trajectories in different regions of the $(w_1,w_2)$ phase-plane under the velocity field $(\dot{w_1},\dot{w_2})$.

Consider the angle $\theta \in [-\pi/2, \pi/2]$ made by the direction $(w_1,w_2)$ with the $w_1$ axis. As before, call $z=w_2/w_1=\tan(\theta)$ and $\beta=qc+(1-q)v$. Then, along a trajectory in the $(w_1,w_2)$ plane,

\begin{eqnarray*}
\dot{z} &=& \frac{d}{dt} \left( \frac{w_2}{w_1} \right) = \frac{\dot{w_2}w_1-\dot{w_1}w_2}{w_1^2}= [(1-q)(v+\delta)+qc]- q\delta z - [(1-q)v+qc]z^2 \\
        &=& -\beta z^2-q \delta z + [\beta+(1-q) \delta]
\end{eqnarray*}

We first want to establish if there are any values of $z$ for which $\dot{z}=0$. These are the slopes along which the rotational speed of the trajectories is zero; in other words, they would correspond to invariant lines in the phase-plane.

We consider the quadratic equation: $\dot{z}=-\beta z^2 - q \delta z + [\beta+(1-q)\delta]=0$. The discriminant is the same as the one of the characteristic equation of ${\bf EC}$:

\begin{eqnarray*}
\Delta &=& q^2 \delta^2+4\beta [\beta+(1-q)\delta] = [2qc+(1-q)(2v_\delta)]^2+(2q-1) \delta^2
\end{eqnarray*}

The solutions of the quadratic equation will the be exactly the slopes of the eigendirections of ${\bf EC}$:

$$z_{1,2}=\frac{-q \delta \pm \sqrt{\Delta}}{2\beta} \in [-\infty,+\infty]$$

proving the following:

\begin{lemma}
The eigendirections of the matrix ${\bf EC}$ represent invariant lines under the vector field of system (1).
\end{lemma}

\begin{figure}[h!]
\begin{center}
\includegraphics[scale=0.37]{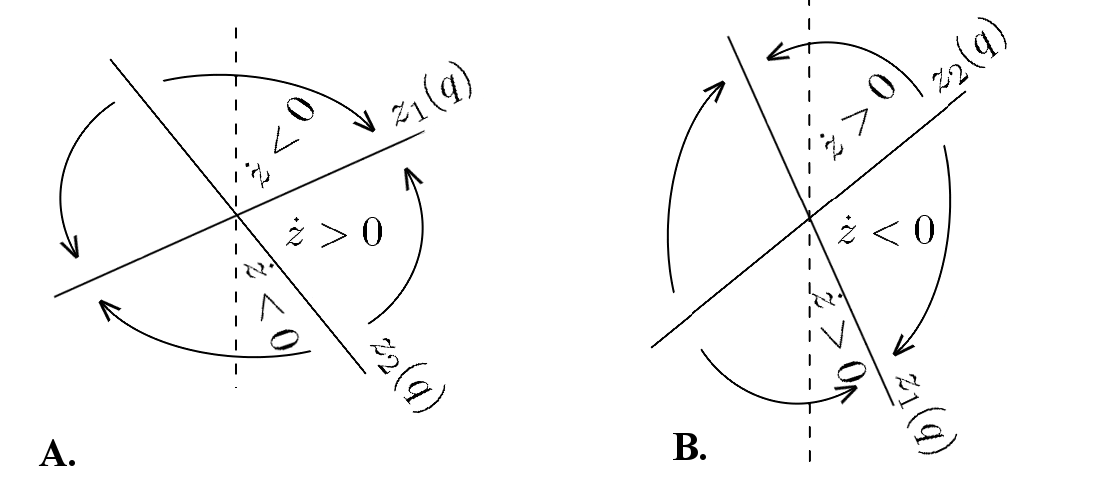}
\end{center}
\caption{{\bf Invariant lines and generic phase plane dynamics.} The invariant lines are marked as $z_1(q)$ and $z_2(q)$. The arrows indicate the rotational direction of the vector field between the two invariant lines. This can be obtained in the right vertical half-plane (where we have defined our angle, $\theta \in [-\pi/2, \pi/2]$), then extended by symmetry in the opposite half-plane. For $q<q^*$ we have $z_1>z_2$ ({\bf A.}). As $q$ increases, the two invariant lines rotate: clockwise if $\delta>0$ and anti-clockwise if $\delta<0$. At $q=q^*$, one of the invariant lines goes through a vertical stage. For $\delta>0$, $\theta_2$ jumps from $-\pi/2$ to $\pi/2$, hence $z_2$ has a vertical asymptote at $q=q^*$, and jumps from $z_2 \to -\infty$ to $z_2 \to \infty$. For $\delta<0$, $\theta_1$ jumps from $\pi/2$ to $-\pi/2$, hence $z_1$ has a vertical asymptote at $q=q^*$, and jumps from $z_1 \to \infty$ to $z_1 \to -\infty$.In consequence, after this critical stage, for $q>q^*$, we have $z_1<z_2$ ({\bf B.}) Although the rotation is continuous, either $z_1$ or $z_2$ has an infinite discontinuity, due to our definition (mod $\pi$) of the angles $\theta_1,2$.}
\label{invlines}
\end{figure}

\vspace{3mm}
We want to better describe the phase-plane behavior \emph{between} the invariant lines $z=z_1$ and $z=z_2$. For any fixed $\displaystyle{q \in (1/2,q^*) \cup (q^*,1]}$ (i.e., for $\beta \neq 0$), the rotational speed is given by the sign of the quadratic function $f(z)=-\beta z^2 - q \delta z + [\beta+(1-q)\delta]$. In principle, we then have two situations:

\begin{enumerate}[i.]
\item $q \in (1/2, q^*)$ (i.e., $\beta >0$). Then $z_1>z_2$, with $z>0$ in $(z_2,z_1)$ and $z<0$ on $(-\infty,z_2) \cup (z_1,\infty)$. The phase plane looks schematically as in Figure~\ref{invlines}a.
\item $q \in (q^*,1]$ (i.e., $\beta <0$). Then $z_1<z_2$, with $z<0$ in $(z_1,z_2)$ and $z>0$ on $(-\infty,z_1) \cup (z_2,\infty)$. The phase plane looks schematically as in Figure~\ref{invlines}b.
\end{enumerate}

\noindent In other words, all trajectories move asymptotically towards the invariant line $z=z_1$.\\

Since the behavior of the system seems to a large extent dictated by these invariant lines, we study how the positions of these lines change under variations of the quality parameter $q$. In other words, we want to study the monotonicity of $z_1=z_1(q)$ and $z_2=z_2(q)$. We get the following (for detailed proofs and limit-case behavior $\displaystyle{\lim_{q \to q^*_{\pm}}{z_{1,2}}}$, see Appendix 3; for illustrations see Figures \ref{invlines} and \ref{phaseplanes}:

\begin{figure}[h!]
\begin{center}
\includegraphics[scale=0.2]{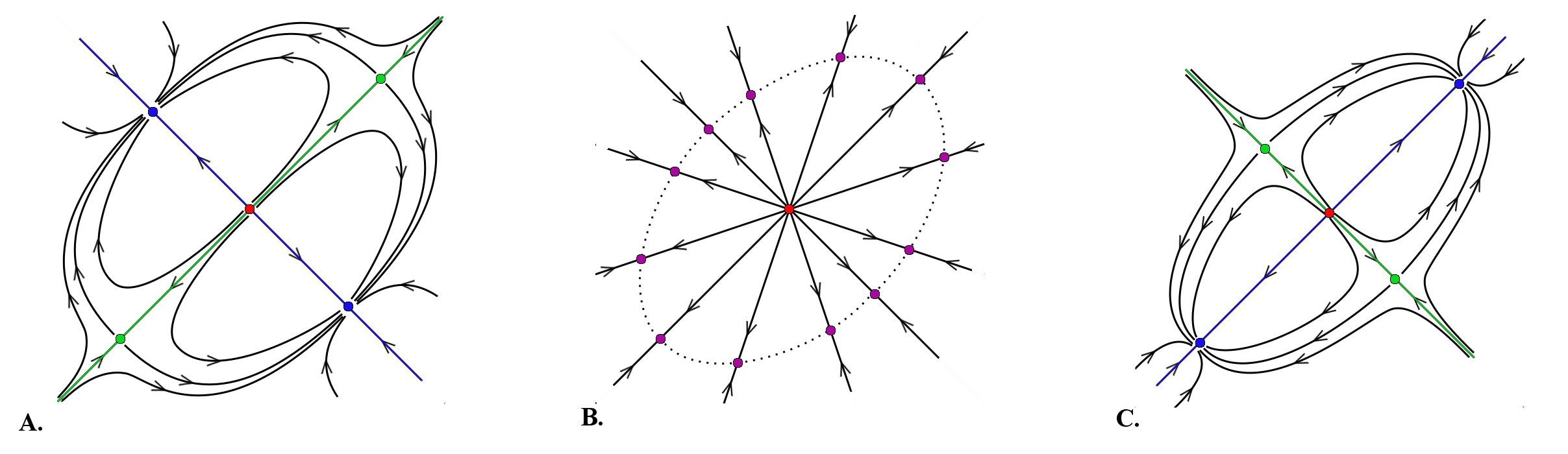}
\end{center}
\caption{{\bf Transitions of the phase plane and bifurcation at $q=q^*$, in the slice $\delta=0$.} {\bf A.} When $q>q^*$, the stable equilibria are the two vectors of norm $\sqrt{q-1/2}$ (blue dots) along the invariant line of slope $z_1=-1$; the saddle equilibria are the two eigenvectors of norm $1$ (green dots) along the invariant line of slope $z_2=1$. As $q$ decreases from $q=1$ towards $q=q^*$, the saddles remain unchanged, but the attractors gradually approach the origin (their norm $\sqrt{q-1/2}$ decreases). {\bf B.} When $q=q^*$, the system traverses a bifurcation state, characterized by an infinite number (an entire ellipse) of neutrally stable equilibria. This critical state permits the swap of stability between the two invariant lines. {\bf C.} When $q<q^*$, the stable equilibria are now the two vectors of norm $1$ (blue dots) along the invariant line of slope $z_1=1$, while the saddle equilibria swapped to the two eigenvectors of norm $sqrt{q-1/2}$ (green dots) along the invariant line of slope $z_2=-1$. As $q$ continues to decrease from $q=q^*$ towards $q=1/2$, the attractors remain unchanged, and the saddles approach the origin (collapsing into the origin in the limit of $q \to 1/2$).}
\label{phaseplanes}
\end{figure}

\begin{prop}
If $\delta<0$, then $\displaystyle{\frac{d z_{1,2}}{dq}>0}$ and hence both $z_1$ and $z_2$ are increasing as $q \in (1/2, q^*) \cup (q^*,1]$. In the system's phase plane, this corresponds to a continuous counter-clockwise rotation of the two invariant lines. If $\delta>0$, then $\displaystyle{\frac{d z_{1,2}}{dq}}<0$; hence both $z_{1,2}$ are in this case decreasing as $q \in (1/2, q^*) \cup (q^*,1]$. In the phase plane, this corresponds to a clockwise rotation of the invariant lines.
\end{prop}

\begin{prop}
The angles $\theta_{1,2} \in [-\pi/2,\pi/2]$ between each invariant line and the $w_1$ abscissa are decreasing with respect to the parameter $q$ in case $\delta>0$, and are increasing with respect to the parameter $q$ in case $\delta<0$. Moreover, in both cases, the angular rate of change is finite, at all $q\in (1/2,1]$.
\end{prop}

\subsection{Unbiased case $\delta=0$}

For $\delta=0$ the computations are simpler; however, as mentioned before, the system has an interesting critical transition which does not appear in the $\delta \neq 0$ slices (occurring from the ``touching,'' or apparent crossing, of the two eigenvalues at $q=q^*$, as shown in Figure~\ref{qvaried}).

\begin{prop}
Suppose $\delta=0$. The phase plane of the system depends on the value of $q$ as follows:
\begin{enumerate}[i.]
\item If $q<q^*$, then $\mu_1=v+c$ is the larger eigenvalue, with eigendirection $z_1=1$ and norm of the corresponding attracting equilibria $\| w \|=1$. $\mu_2=(2q-1)(v-c)$ is the smaller eigenvalue, with eigendirection $z_2=-1$ and norm of the corresponding saddle equilibria $\| w \|=\sqrt{q-1/2}$.
\item If $q>q^*$, then $\mu_1=(2q-1)(v-c)$ is the larger eigenvalue, with eigendirection $z_1=-1$ and norm of the corresponding attracting equilibria $\| w \|=\sqrt{q-1/2}$. $\mu_2=v+c$ is the smaller eigenvalue, with eigendirection $z_2=1$ and norm of the corresponding saddle equilibria $\| w \|=1$.
\item If $q=q^*$, the system contains an infinity of half-stable non-isolated equilibria (each direction will contain two opposite equilibria, describing overall an ellipse of equilibria around the origin).\\
\end{enumerate}
\end{prop}

\proof{For $\delta=0$, we have $\dot{z}=-\beta(z^2-1)$. The situation $q<q^*$ corresponds to $\beta>0$, and $q>q^*$ corresponds to $\beta<0$. Parts \emph{i.} and \emph{ii.} follow immediately. For $q=q^*$, $\dot{z}=0$; all lines through the origin are invariant, and each contains two half-stable equilibria. In the Appendix 4, we show that the locus of these equilibria is an ellipse (see dotted curve in Figure~\ref{phaseplanes}), and we describe its axes and foci.}

\begin{figure}[h!]
\begin{center}
\includegraphics[scale=0.25]{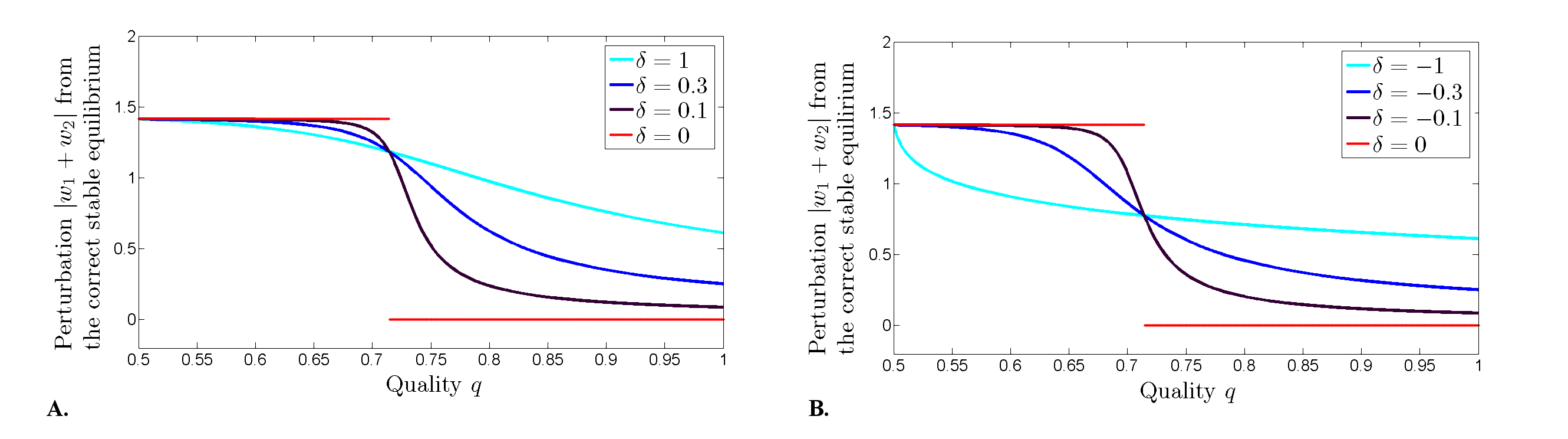}
\end{center}
\caption{{\bf $S(q)=\lvert w_1+w_2 \rvert$ as a measure of the increasing inspecificity of the stable equilibrium, compared to its ideal state $S(1)=0$, as $q$ decays from $q=1$.} For $v=1$, $c=-0.4$, we plotted $S(q)$. {\bf A.} For $\delta \geq 0$: $\delta=1$ (cyan); $\delta=0.3$ (blue); $\delta=0.1$ (purple); $\delta=0$ (red). {\bf B.} For $\delta \leq 0$: $\delta=-1$ (cyan); $\delta=-0.3$ (blue); $\delta=-0.1$ (purple); $\delta=0$ (red). In both panels, all continuous curves for $\delta \neq 0$ concur at one point, which corresponds to the fact that, for both $\delta>0$ and $\delta<0$, the stable equilibrium at $q=q^*$ is independent on the magnitude of $\delta$.}
\label{inspecific}
\end{figure}

\vspace{3mm}
\noindent {\bf Remark 1.} For $q=1$, the attracting equilibria lay along the direction $z=-1$, so that $w_1+w_2=0$. A simple way to quantify how far the stable equilibrium $w=(w_1,w_2)$ degrades from this error-free state as the quality $q$ decreases, we can measure how much the sum $S(q)=\lvert w_1+w_2 \rvert$ deviates from zero, the outcome of perfect learning (Figure~\ref{inspecific}).

In Figure~\ref{qvaried}b, the inputs are unbiased ($\delta=0$), and in the absence of crosstalk ($q=1$) the inputs segregate completely. As crosstalk increases, the separation between the eigenvalues at first decreases, though the inputs remain completely segregated. However, as crosstalk increases further, the two eigenvalues equalize at the critical quality value $q^*=v/(v-c)$. With further increases in crosstalk, the inputs become completely unsegregated, and the eigenvalues now move apart. This qualitative change at $q^*$ is a bifurcation. Note that although the qualitative behavior only changes at $q^*$, there is a biologically less important quantitative change: the two symmetric equilibrium weight vectors decrease continuously in length as $\sqrt{q-1/2}$, as $q$ decreases from $q=1$ until the bifurcation at $q=q^*$, then remain of unit length for $q<q^*$.

\begin{figure}[h!]
\begin{center}
\includegraphics[scale=0.35]{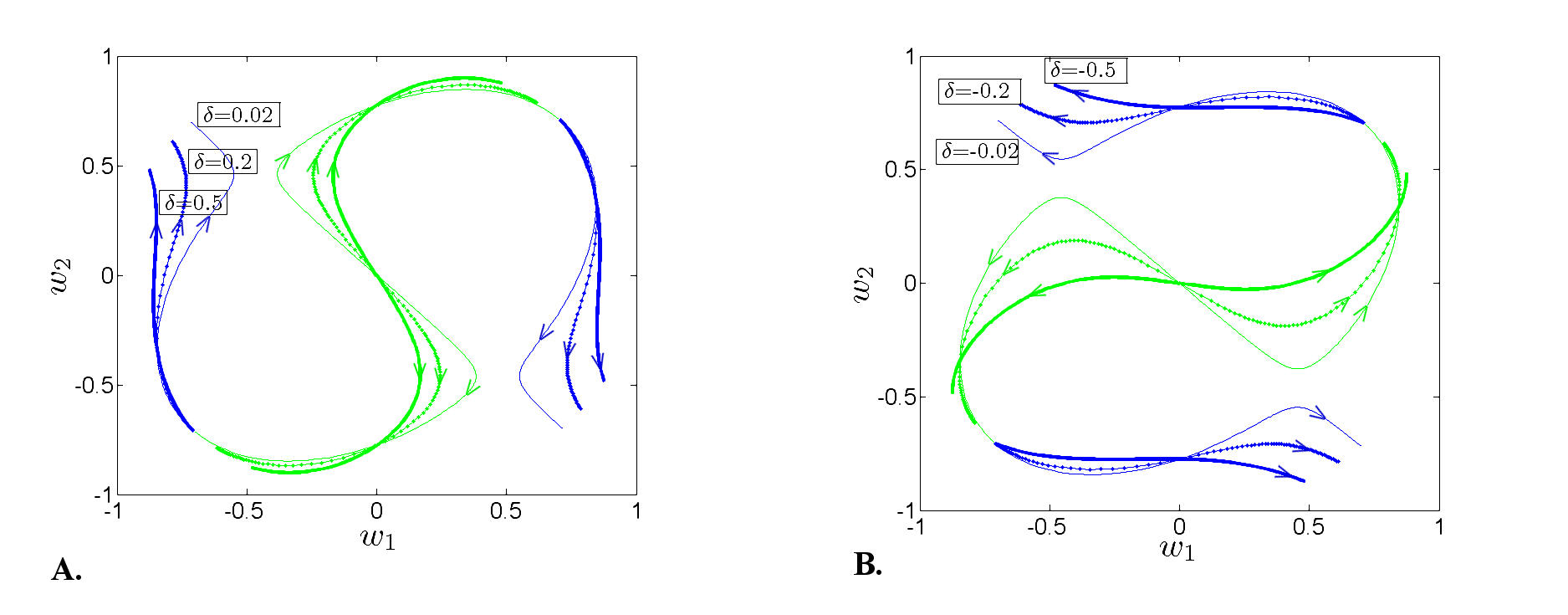}
\end{center}
\caption{{\bf Equilibria curves in the phase plane, as $q$ changes.} The blue curves represent the stable equilibrium locus, and the green curves the saddle equilibrium. {\bf A.} Plots for a few representative positive $\delta$ values: $\delta=0.02$ (thin curves), $\delta=0.2$ (thin dotted curves) and $\delta=0.5$ (thick curves). All green saddle curves concur at one point (on the vertical axis), and all blue stable curves also concur at a point, corresponding to the fact that the position of the two equilibria is independent on the magnitude of $\delta>0$. {\bf B.} Plots for a few representative negative $\delta$ values: $\delta=-0.02$ (thin curves), $\delta=-0.2$ (thin dotted curves) and $\delta=-0.5$ (thick curves). All green saddle curves concur at one point, and all blue stable curves also concur at a point (on the vertical axis), corresponding to the fact that the position of the two equilibria is independent on the magnitude of $\delta<0$. The arrows along the curves indicate the direction of increasing $q$.}
\label{equilibria}
\end{figure}

In the slightly unbiased cases $\delta=-0.2$ and $0.5$ (Figure~\ref{qvaried}a and c), this overall behavior persists, although the eigenvalues always remain distinct, and there is no true bifurcation. Thus in part A, as crosstalk increases, the eigenvalues at first approach each other, and the solution remains almost segregated. At the ``pseudocritical'', value of $q=\frac{(2v+\delta)(2v+\delta-2c)-\delta^2}{2v+\delta-2c}$, the eigenvalues reach their closest value, (in an ``avoided crossing'') and then start to separate as crosstalk increases further; significantly beyond this pseudocritical value, the outcome is almost unsegregated (see Figures~\ref{inspecific} and ~\ref{equilibria}). Of course, for $q$ values very close to this pseudocritical value, desegregation is very rapidly increasing with increases in crosstalk (see Figures~\ref{inspecific} and ~\ref{equilibria}), especially with very small values of $\delta$. Thus even with slight input bias, the overall behavior, switching from segregation to unsegregation at a critical crosstalk value, resembles that seen in the unbiased situation.

\subsection{Conclusions: mathematical behavior of the 2D system}

\begin{corol}
For any combination of parameters, the phase-plane of the system (1) contains no cycles. Moreover, the system has only two (opposite) attracting equilibria, with attraction basins two open half-planes.
\end{corol}

\noindent {\bf Remark.} The result holds more generally for an n-dimensional system, as shown in Appendix 1.

\vspace{3mm}
Since we are looking at a 2-dimensional system, this means, according to the Poincar\'{e}-Bendixon theorem, that the only attracting sets can be attracting equilibria. The two attracting equilibria of the system (by Proposition 2.3) lie along the invariant line corresponding to the largest eigenvalue of the covariance matrix ${\bf C}$, hence their position (direction and distance to origin) depend on the values of the parameters (in particular on the quality $q$ and bias factor $\delta$.). Figure~\ref{equilibria} illustrates the evolution of these points in the phase plane for a fixed $\delta \neq 0$, as $q$ increases. (We used Matcont continuation algorithms to numerically estimate the equilibria and draw the equilibrium curves.)

The following two paragraphs summarize the conclusions obtained throughout the previous sections:

\vspace{3mm}
{\bf Biased dynamics.} \emph{When the system is biased (i.e., $\delta \neq 0$) the two eigenvalues of the input covariance matrix ${\bf C}$ are always separated. The phase plane has two pairs of nonzero opposing equilibria, each situated on one of two distinct invariant lines through the origin (i.e., the two eigendirections of ${\bf EC}$). The invariant line of slope $z_1$ corresponding to the higher eigenvalue $\mu_1$ of ${\bf EC}$ contains the pair of opposing attracting equilibria; the invariant line of slope $z_2$ corresponding to the lower eigenvalue $\mu_2$ of ${\bf EC}$ separates their two basins of attraction and also contains the pair of opposing saddles. As the parameter $q$ increases, the invariant lines rotate (clockwise if $\delta>0$ and counter-clockwise if $\delta<0$) in a continuously differentiable manner, with an angular speed that depends on $q$. This rotation gets arbitrarily fast (e.g., at its point of maximal rotational speed) as $\delta \to 0$.}

\vspace{3mm}
{\bf Unbiased dynamics.} \emph{When the system is unbiased (i.e., $\delta = 0$) the two eigenvalues of the input covariance matrix ${\bf C}$ collide at the critical value of the quality parameter $q=q^*$. For any $q \neq q^*$, the phase plane has two pairs of nonzero opposing equilibria, each situated on one of two distinct invariant lines $\left( \begin{array}{r} 1 \\ \pm 1 \end{array} \right)$ through the origin. The invariant line corresponding to the higher eigenvalue of ${\bf EC}$ contains the pair of opposing attracting equilibria; the invariant line corresponding to the lower eigenvalue of ${\bf EC}$ separates their two basins of attraction and also contains the pair of opposing saddles. As the parameter $q$ increases, the invariant lines remain unchanged, until they swap instantaneously as $q$ traverses the critical state $q=q^*$ (stability-swapping {\bf bifurcation}). At the bifurcation point, the phase plane has an entire ellipse of half-stable equilibria.}

\begin{figure}[h!]
\begin{center}
\includegraphics[scale=0.25]{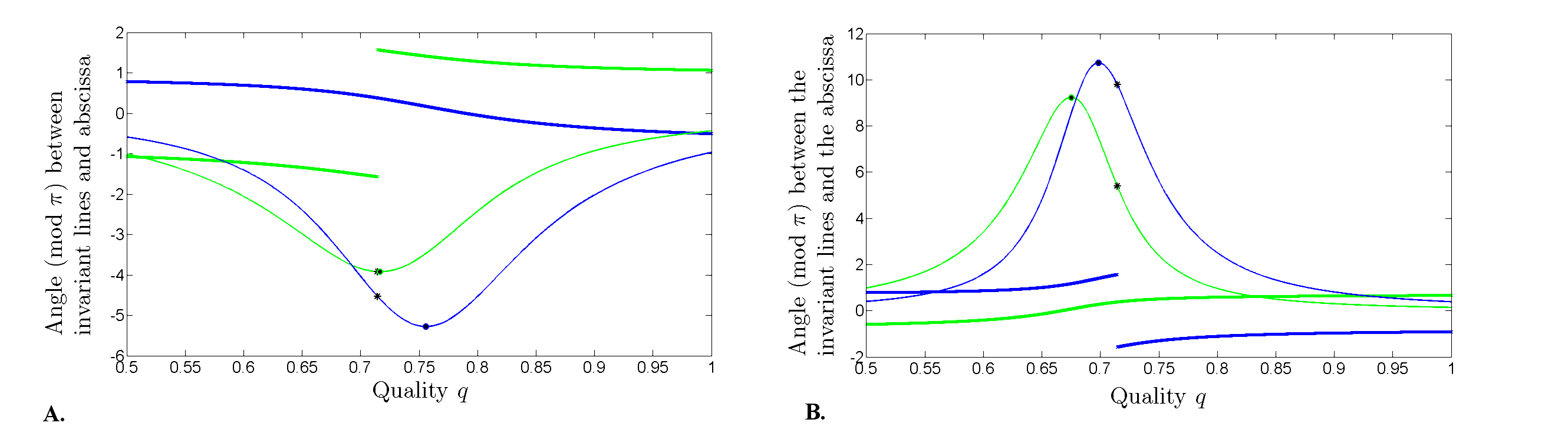}
\end{center}
\caption{{\bf Illustration of the evolution of the angles $\theta_{1,2}$ of the invariant lines with the abscissa, as $q$ increases.} In both panels, $v=1$ and $c=-0.4$. {\bf A.} $\delta=0.5$; {\bf B.} $\delta=-0.2$. The graphs of the functions are shown in thick lines, $\theta_1$ in blue and $\theta_2$ in green. The graphs of the derivatives are plotted in thin lines, with $d \theta_1/dq$ in blue and $d \theta_2/dq$ in green. On the graphs of the derivatives, we marked with a black star the points corresponding to $q=q^*$, and with a bullet the points of extremum (the inflection points for $\theta_{1,2}$, where the rotational speed is maximal).}
\label{angles}
\end{figure}

\vspace{3mm}
\noindent {\bf Remark.} The {\bf codimension 2 bifurcation} that occurs at $q=q^*$ in the slice $\delta=0$ can be considered a limit case of the phase-plane transition sequence obtained when increasing $q$, when making $\delta \to 0$ in the biased case. The rotational speed blows up to $\infty$ as $\delta \to 0$, and, in the $\delta=0$ slice, the rotation becomes instantaneous via what appears to be the bifurcation's ``swap'' of eigendirections. The evolution of the rotation speed with respect to $q$ as $\delta \to 0$ is further illustrated in Figure~\ref{angles}.

\section{Alternative models: Euclidean normalization of weights versus the Oja model}

The Oja rule is an elegant and classical solution to the well-known problem that unconstrained Hebbian learning is unstable~\cite{dayan2001theoretical,oja1982simplified}. It has the biologically appealing feature that it is local, although it does require, somewhat implausibly, that the ``normalizing'' adjustment is proportional to the current weight. We have shown that it is still useful when some crosstalk is present, although the stable norm, and the exact direction of the learned weight vector, changes. One can imagine various other ways, possibly involving ``homeostasis'' or ``synaptic scaling''~\cite{turrigiano1998activity,turrigiano2004homeostatic} of promoting stability, and some studies invoke various combinations of these mechanisms. A less biologically plausible, nonlocal but extremely simple and highly effective method, which might capture features of any more plausible scheme and which works even for nonlinear rules, is  to impose a specific norm after each weight vector update. Here we examine how crosstalk affects such ``explicit'' or ``brute'' normalization.\\

\noindent As before, Hebb's rule lies at the basis of the weight updates: $\Delta {\bf w} = \gamma y{\bf x}$, with $y={\bf w}^t {\bf x}={\bf x}^t {\bf w}$.\\

\noindent In other words: ${\bf w}(n+1) = {\bf w} + \gamma y{\bf x}$. As in the Oja model, we can think of Hebbian inspecificity being formalized as a stochastic error matrix \mbox{\boldmath ${\cal E}$}, so that, at each time step:

$${\bf w} \to {\bf w} + \gamma y{\bf\mbox{\boldmath ${\cal E}$}x}$$

\noindent Taking expectation of both sides and re-naming ${\bf w}=\langle {\bf w} \rangle$ (the long-term average of the weight vector), ${\bf C}=\langle {\bf x}^t{\bf x} \rangle$ (the correlation matrix of the input distribution) and ${\bf E}=\langle $\mbox{\boldmath ${\cal E}$}$\rangle$ (the average error matrix), we obtain the iteration: ${\bf w} \to {\bf w} + \gamma \langle$\mbox{\boldmath ${\cal E}$}${\bf xx}^t \rangle {\bf w} = {\bf w} + \gamma {\bf ECw}$.\\

\noindent We normalize to keep $\| {\bf w} \| = 1$, and make no further approximations to implement this normalization biologically. We get the new iteration function that describes the average iterative process, with errors, becomes:

$$f({\bf w}) = \frac{{\bf w}+ \gamma {\bf ECw}}{\| {\bf w} + \gamma {\bf ECw} \|}$$

\noindent where the ``modified'' covariance matrix is as before ${\bf EC}$; unlike in the Oja case, ${\bf EC}$ is now involved in the normalization step as well. Notice that, since ${\bf EC}$ has positive eigenvalues, the matrix ${\bf I}+\gamma {\bf EC}$ is nonsingular, hence $f$ is defined for all ${\bf w} \in {\mathbb R}^n \backslash \{0\}$. The rest of the section is dedicated to discussing the position and stability of the equilibria of this new system, in whose case the direct normalization confines the trajectories to the unit circle.

\vspace{3mm}
In order to slightly simplify the notation, we call ${\bf A}={\bf EC}$, ${\bf u}={\bf w}+\gamma {\bf ECw}$ and $a=\| {\bf u} \|$, notation which we will use whenever it is convenient. We want to see if the long-term evolution of ${\bf w}$ predicted by this model is comparable with the behavior of our stochastic, discrete simulations for case where the ``ratio'' normalization was replaced by a ``subtractive'' Taylor approximation of it (see also ~\cite{radulescu2009hebbian}.\\

\vspace{5mm}
\noindent The vector ${\bf w}$ is a fixed point of $f({\bf w})$ iff ${\bf w} + \gamma {\bf Aw} = a{\bf w}$, i.e. ${\bf w}$ is a unit eigenvector of ${\bf A}$ (with the Euclidean norm). To establish the stability, we compute the Jacobian matrix of $f$ at each fixed point.

\vspace{3mm}
Fix $j \in \overline{1,n}$. Then, for any $i \neq j$:

\begin{equation}
\frac{\partial u_i}{\partial w_j} = \frac{\partial}{\partial w_j}{(w_i+\gamma [{\bf Aw}]_i)} = \gamma A_{ij} \nonumber
\end{equation}

\noindent  When $i=j$, we have similarly:

\begin{equation}
\frac{\partial u_j}{\partial w_j} = \frac{\partial}{\partial w_j}{(w_j+\gamma [{\bf Aw}]_j)} = 1 + \gamma A_{jj} \nonumber
\end{equation}

\noindent Hence, overall:

\begin{eqnarray*}
\frac{\partial}{\partial w_j} \| {\bf u} \| ^2 &=& 2 u_j (1+\gamma A_{jj}) + \sum_{i \neq j} 2 u_i \gamma A_{ij} = 2u_j + 2\gamma \sum_i{u_i A_{ij}} = 2u_j+2\gamma [{\bf A}^t{\bf u}]_j
\end{eqnarray*}

\noindent In matrix form:

\begin{equation}
\frac{\partial}{\partial {\bf w}} \| {\bf u} \| ^2 = 2\gamma {\bf A}^t{\bf u} + 2{\bf u}
\end{equation}

\vspace{3mm}
\noindent Now, fix $i \in \overline{1,n}$. For $j \neq i$, we have:

\begin{equation}
\frac{\partial f_i}{\partial w_j} = \frac{\gamma A_{ij} \|u\| - u_i \| {\bf u} \| ^{-1} [\gamma {\bf A}^t{\bf u} + {\bf u}]_j}{\| {\bf u} \| ^2}  \nonumber
\end{equation}

\noindent For $j=i$, we have:

\begin{equation}
\frac{\partial f_i}{\partial w_i} =  \frac{(1 + \gamma A_{ii}) \| {\bf u} \| - u_i \| {\bf u} \| ^{-1} [\gamma {\bf A}^t{\bf u} + {\bf u}]_i}{\| {\bf u} \| ^2}  \nonumber
\end{equation}

Rewritten in matrix form:

\begin{equation}
\frac{\partial f}{\partial {\bf w}} = \frac{\gamma}{a}{\bf A} - \frac{1}{a^3} (\gamma {\bf uu}^t{\bf A} + {\bf uu}^t) +  \frac{1}{a} {\bf I} = \frac{1}{a}\left( {\bf I}-\frac{1}{a^2}{\bf uu}^t \right)(\gamma {\bf A} + {\bf I})
\end{equation}

\noindent where ${\bf I}$ is the appropriate size identity matrix.\\

At any fixed point ${\bf w}$, for which automatically $\| {\bf w} \| = 1$ and ${\bf Aw}=\lambda_{\bf w} {\bf w}$, where $1 + \lambda_{\bf w} \gamma=a$), we have that:

\begin{eqnarray*}
{\bf uu}^t &=& ({\bf w} + \gamma {\bf Aw}) ({\bf w} + \gamma {\bf Aw})^t = (1 + 2 \gamma \lambda_{\bf w} + \gamma^2 \lambda_{\bf w}^2) {\bf ww}^t = (1 + \lambda_{\bf w} \gamma)^2 {\bf ww}^t
\end{eqnarray*}

\noindent The Jacobian at a fixed point ${\bf w}$ can be then simplified to:

\begin{equation}
\frac{\partial f}{\partial {\bf w}} = \frac{1}{a}({\bf I} - {\bf ww}^t)(\gamma {\bf A} + {\bf I})
\end{equation}

\noindent We calculate:

\begin{eqnarray*}
\frac{\partial f}{\partial {\bf w}}({\bf w}) &=& \frac{1}{a}({\bf I}-{\bf ww}^t)(\gamma \lambda_{\bf w} + 1){\bf w} = \frac{1}{a}({\bf w}-{\bf w}({\bf w}^t{\bf w}))(\gamma \lambda_{\bf w} + 1) = 0
\end{eqnarray*}

Complete ${\bf w}$ to a basis of eigenvectors of ${\bf A}$ (not necessarily mutually orthogonal). Let ${\bf v} \neq {\bf w}$ any of the vectors in this basis (with eigenvalue $\lambda_{\bf v}$), and consider ${\bf z} = {\bf v}- [{\bf w}^t{\bf v}]{\bf w}$ the projection of ${\bf v}$ on the orthogonal complement of ${\bf w}$. Then:

\begin{eqnarray*}
(\gamma {\bf A} + {\bf I})({\bf z}) = (\gamma \lambda_{\bf v} + 1){\bf v} - (\gamma \lambda_{\bf w} + 1)[{\bf w}^t{\bf v}]{\bf w}
\end{eqnarray*}

\noindent Hence

\begin{eqnarray*}
\frac{\partial f}{\partial {\bf w}}({\bf z}) = \frac{1}{a} (\gamma \lambda_{\bf v} + 1)({\bf v} - [{\bf w}^t{\bf v}]{\bf w}) = \frac{1}{a} (\gamma \lambda_{\bf v} + 1){\bf z} = \frac{\gamma \lambda_{\bf v} + 1}{\gamma \lambda_{\bf w} + 1} \, {\bf z}
\end{eqnarray*}

\noindent A normalized eigenvector ${\bf w}$ of ${\bf EC}$ is stable as a fixed point of the system if all the eigenvalues of the Jacobian $\frac{\partial f}{\partial {\bf w}}$ at ${\bf w}$ are less than one in absolute value :

$$\left \lvert \frac{\gamma \lambda_{\bf v} + 1}{\gamma \lambda_{\bf w} +1} \right \rvert < 1 $$

\noindent Since all eigenvalues of ${\bf EC}$ are positive (recall that ${\bf EC}$ is diagonalizable with the dot product $\langle \cdot , \cdot \rangle_{\bf C}$), this is equivalent to $\ds \frac{\gamma \lambda_{\bf v} + 1}{\gamma \lambda_{\bf w} +1} < 1$, and thus to $\lambda_{\bf w} > \lambda_{\bf v}$ for every ${\bf v} \neq {\bf w}$.

{\bf In conclusion:} The system has stable fixed points iff the modified correlation matrix ${\bf EC}$ has a maximal eigenvalue of multiplicity one. Then, a point ${\bf w}$ is a stable fixed point of the system iff it is a unit eigenvector of ${\bf EC}$ corresponding to the unique maximal eigenvalue of ${\bf EC}$''.

It is now clear that the phase space of this system, although not dynamically equivalent to the phase space of the corresponding Oja model, it is very similar. Disregarding the origin (which is not in the domain of one, but is a repelling fixed point for the other), the other fixed points have the same qualitative behavior (stability) for both systems, if assuming $\gamma$ sufficiently small. Moreover, the stability transitions occur at the same bifurcation points (where the eigenvalues of ${bf EC}$ collide with each other), and the bifurcation phase-planes are themselves similar.\\

In the case of the two-dimensional model discussed in this paper, the eigenvalue swap occurs as before when $\ds q=q^*=\frac{v}{v-c}$. For example, in the unbiased case $\delta=0$, the bifurcation phase plane at $q=q^*$ again exhibits an ellipse attractor. Indeed, at the codimension 2 $q=q^*$, the iteration function becomes: $\ds f({\bf w}) = \frac{{\bf w}}{\| {\bf w} \|}$, which maps radially any ${\bf w}$ in the plane to the unit circle, and keeps it fixed thereafter.

The next section shows phase plane simulations for both models, in the more realistic situation of stochastic weight updates, in discrete time and at at finite learning rate.

\section{Stochastic models. Simulations and predictions}

Here we briefly study the more biologically realistic situation in which the weights update stochastically with a small finite learning rate $\gamma$, driven by each individual input, rather than by the mean statistics in the negligible learning rate limit. In particular we study whether the convergence to eigenvector equilibria, and the transitions in dynamics between different values of the parameter $q$ still occur as in the deterministic model.

\begin{figure}[h!]
\begin{center}
\includegraphics[scale=.17]{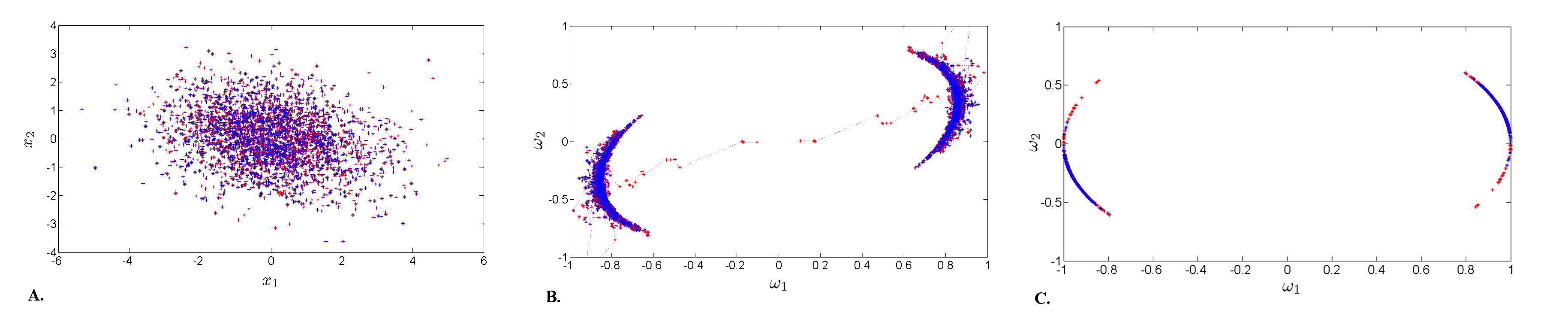}
\end{center}
\caption{{\bf Behavior of stochastic weight updates for a biased input distribution $\delta \neq 0$.} {\bf A.} A discrete input sample ($N=4000$) was drawn out of an input distribution with $v=1$, $c=-0.4$, $\delta=1$, and used to update the weights. {\bf B.} Depending on their initial state, the weight vector stabilizes towards small stochastic fluctuations around either one of the attracting equilibria (the pair of appropriately normalized eigenvectors corresponding to the larger eigenvalue of ${\bf EC}$). {\bf C.} The corresponding iterations are shown in the case of exact normalization at each step (fewer iterations are shown in this case, since more weights, all living on the unit circle, would obstruct the clarity of the figure.) In all three panels, the points were colored update-chronologically from red to blue. We used the critical quality $q=1/1.4 \sim 0.71$.}
\label{simbiased}
\end{figure}

\begin{figure}[h!]
\begin{center}
\includegraphics[scale=.25]{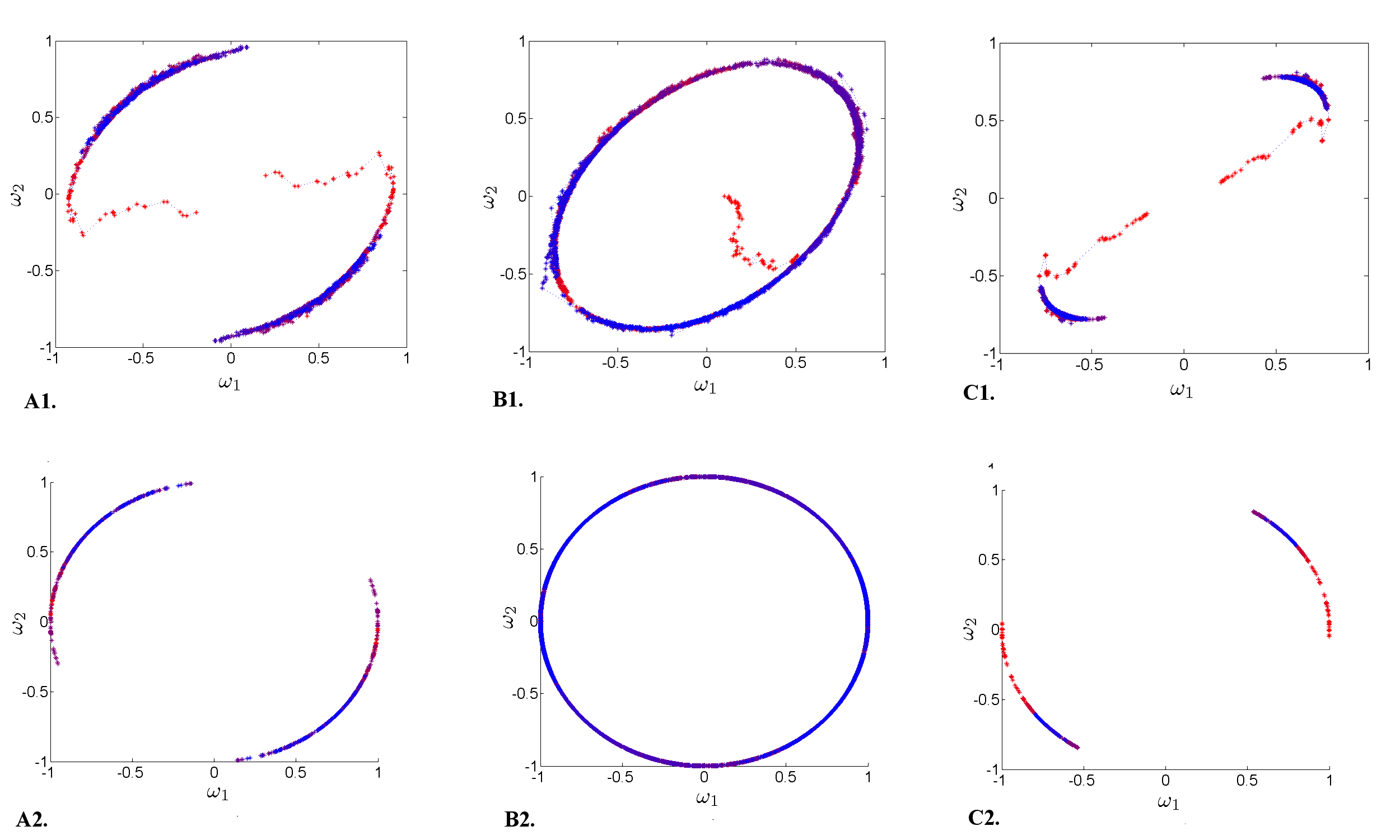}
\end{center}
\caption{{\bf Differences in stochastic behavior when $q$ is varied, in the unbiased input case $\delta=0$ (compare with Figure~\ref{phaseplanes}).} A discrete number of input vectors ${\bf x}(t)=(x_1(t),x_2(t))$ are drawn from a distribution with covariance matrix $C$, with $v=1$, $c=-0.4$ (so that the critical quality value $q^*=1/1.4 \sim 0.71$). The weights \mbox{\boldmath $\omega$}$=(\omega_1(t),\omega_2(t))$, adjusting with a small learning rate $\gamma=0.1$, are plotted in the $(\omega_1,\omega_2)$ plane, with the color of the points changing chronologically from red to blue. The top panels show the behavior of the Oja model, while the bottom panels, for corresponding parameters, show the behavior for exact normalization of weights at each step. {\bf A.} For good transmission quality $q=0.85>q^*$, \mbox{\boldmath $\omega$} is converging in the long term to a state of small fluctuations around either $\displaystyle{\pm \sqrt{q-1/2} \, (1 \; -1)^T}$, depending on the initial state. The plot illustrates the trajectories for two initial states, each stabilizing around one of these opposite eigenvectors. {\bf B.} For critical transmission quality $q=q^*$, \mbox{\boldmath $\omega$} converges to fluctuations around the ellipse of neutrally attracting equilibria, but will perpetually drift around, filling the ellipse, driven by input fluctuations from the mean statistics, without remaining asymptotically near any particular equilibrium state. {\bf C.} For poor transmission quality $q=0.6<q^*$, \mbox{\boldmath $\omega$} is converging in the long term to a state of small fluctuations around $\displaystyle{\pm \frac{1}{\sqrt{2}} \, (1 \; 1)^T}$, depending on the initial state. The plot illustrates the trajectories for two initial states, each stabilizing to stabilizing around one of these opposite eigenvectors.}
\label{simunbiased}
\end{figure}

Our numerical simulations show, as expected, that convergence is conserved, in the following sense: when a pair of attracting equilibria exist for the deterministic system (i.e., ${\bf EC}$ has distinct eigenvalues), the discrete sequence of updating \mbox{\boldmath $\omega$} eventually stabilizes to small, stochastic fluctuations around one of these two equilibria (which are, as we recall, the appropriately normalized eigenvectors corresponding to the larger eigenvalue of ${\bf EC}$). This is illustrated in Figures 8, 9a and 9c. In Figure~\ref{simbiased}, ${\bf x}$ is drawn out of a biased distribution of inputs (shown on the left), for which the two eigenvalues of ${\bf EC}$ are warranted to be distinct for any value of $q$, in particular for the value chosen here ($q=q^*=1/1.4$). In Figure~\ref{simunbiased}, the inputs are unbiased, so the same remark applies only if $q \neq q^*$. In Figure~\ref{simbiased}a, we illustrate the case $q>q^*$, in which the attracting vectors are $\displaystyle{\pm \sqrt{q-1/2} \left( \begin{array}{r} 1 \\ -1 \end{array} \right)}$; in Figure~\ref{simbiased}c, we illustrate the case $q<q^*$, in which the attracting vectors are $\displaystyle{\pm \frac{1}{\sqrt{2}} \left( \begin{array}{r} 1 \\ 1 \end{array} \right)}$. In both cases, the stochastic update settles to fluctuations about either one of these vectors, depending on the initial conditions.

Figure~\ref{simunbiased}b illustrates the unbiased case corresponding to the codimension 2 bifurcation in the deterministic dynamics; that is, when $q=q^*$. Recall that, in the deterministic phase-plane, this case was characterized by an ellipse of neutrally stable equilibria, so that each initial condition would converge radially towards a unique nonisolated equilibrium on this curve. This situation changes in the model driven by stochastic updates. An initial weight vector \mbox{\boldmath $\omega$} will quickly be attracted towards the ellipse; however, the orbit does not fluctuate around a particular point on the curve, but rather perpetually drifts along the curve, eventually covering densely the entire ellipse.

\section{Discussion}

We have proposed~\cite{cox2009hebbian,adams2002new} that a central problem for biological learning is that the activity-dependent processes that lead to connection strength adjustments cannot be completely synapse specific, because they must obey the laws of physics. This truism provides a new viewpoint: it raises the possibility that sophisticated learning, such as presumably occurs in the neocortex, is enabled as much by special machinery for enhancing specificity, as by special algorithms~\cite{adams2006neurobiological}. We have suggested that these plasticity errors are analogous to mutations, and that cortical circuitry might reduce such errors, just as ``proofreading'' reduces dna copying mistakes. In particular, it seems possible that the key to overcoming the curse of dimensionality that underlies difficult, and apparently almost intractable, learning problems lies not just in finding good approximations, architectures and techniques, but also in perfecting the relevant biological plasticity apparatus. Indeed, it seems possible that problems of survival and reproduction are so diverse that no single algorithm can solve them all, so that no ``universal'' or ``canonical'' cortical circuit would be expected. In these circumstances, as Rutherford once said about physics, neuroscience would become a type of stamp collecting. However, if every specialized algorithm relies on extraordinarily specific synaptic weight adjustment, then finding machinery that allows such specificity would indeed be tantamount to discovering new neurobiological general principles, somewhat along the lines that established the main framework for modern biology (Darwinian evolution, Mendelian genetics, DNA structure and function, replication mechanisms etc).  We have speculated that an important part of such machinery, at least in the neocortex, might lie outside the synapse itself, in the form of complex circuitry performing a proofreading operation analogous to that procuring accuracy for polynucleotide copying~\cite{adams2002new,adams2006neurobiological,biomathics}. However, such machinery would be less necessary if update inaccuracy merely degraded learning, rather than preventing it. In particular even if temporarily unfavorable (e.g., ``noisy'') input statistics led to imperfect learning because of Hebbian inspecificity, the degraded weights might still be a useful starting point for better learning when input statistics improve. On the other hand, if inspecificity completely prevented even partial learning, then rapid and successful learning from newly favorable statistics might be impossible. These considerations have impelled us to examine the effect of Hebbian ``crosstalk'' in various classical models of unsupervised learning, using both linear~\cite{radulescu2009hebbian} and nonlinear rules~\cite{cox2009hebbian} (see also ~\cite{elliott2012}).

\subsection{Separate but equal: segregation without bias}

In this paper we extended our previous study~\cite{radulescu2009hebbian} of the effect of crosstalk on the simple linear Hebbian model of Oja to situations approaching the ``unbiased'' case where all inputs have the same statistical distribution. This case has often been invoked in discussions of the emergence of ocular dominance wiring and other forms of neural development, but it might also apply to any situation in which sets of inputs disconnect completely, or ``segregate,'' to form pruned wiring patterns that are then ``sculpted'' by a more subtle synaptic learning process (of course in the present model weights and activities can be negative; we interpret negative weights as disconnections). For the case of visual input, it seems likely that statistics would be similar, and positively correlated, for the two eyes, which look at the same world, and it is well known~\cite{dayan2001theoretical} that a linear Hebb rule with unbiased inputs, under either implicit or explicit normalization, leads to the symmetric, equal-weight, and thus apparently unbiological, outcome. A possible but rather unbiological solution to this is to use a ``subtractive'' normalization scheme, although this also requires imposing weight limits~\cite{miller1994role}. It has been shown that a wide variety of nonlinear rules~\cite{elliott2003analysis}, including the BCM rule~\cite{bienenstock1982theory} and STDP~\cite{elliott2008temporal} can lead to ocular segregation under unbiased statistics. The key point is that segregated states can be created by typically nonlinear, ``symmetry-breaking'' mechanisms even when the inputs themselves do not favor particular segregated outcomes. Indeed, the absence of bias could be characteristic of development, as opposed to ``learning,'' insofar as these two notions are distinct.

A natural question would be: if such segregated outcomes are an important part of normal development (which then constrain subsequent, more detailed, ``refining,'' plasticity processes, including learning), how could the determining ``unbiased'' statistics arise, and conversely, how would plasticity errors, such as crosstalk, or other alterations in the form of the rule, affect the outcome? In particular we show here that, unsurprisingly, crosstalk tends to prevent segregation, especially when the inputs are close to unbiased. This might set a limit to the use of symmetry breaking to generate specific wiring, or require special specificity-enhancing circuitry, such as "proofreading", even during development. At the very least it suggests that internally generated patterns deriving segregation, such as negative correlations, might have to be quite strong to overcome the desegregating effect of inevitable crosstalk.

 Before exploring this further, we comment briefly about ``unbias'' in relation to Hebbian learning. Although here we focus on lack of bias in the second order statistics, one can also postulate unbias at all order, an assumption which greatly simplifies the study of nonlinear Hebbian plasticity, essentially eliminating the possibility of learning and restricting analysis to development. To what would unbiased high-order statistics correspond? It seems that they correspond to the radially symmetric distributions recently considered by Lyu and Simoncelli~\cite{lyu2009nonlinear}, where the joint pdf equal density contour lines are nested hyperspheres  with nonGaussian spacings. One might expect that with completely unbiased (spherical) input statistics no particular direction in weight space would be favored and therefore the outcomes would be either symmetric (equal weights), or broken symmetric (various combinations of opposite but equal magnitude weights); the particular set of outcomes would be determined by the higher-order correlations, and could be quite complicated. Indeed, Elliott~\cite{elliott2003analysis} finds that segregated outcomes are quite typical of nonlinear Hebbian rules with unbiased statistics and shows that crosstalk can induce bifurcations in these cases~\cite{elliott2012}.

 Recently, it has been suggested that the Oja rule (even without crosstalk) and Eigen's replication/mutation equation might be ``isomorphic''~\cite{fernando2009chemical,fernando2010neuronal}. Indeed both equations describe normalized growth processes. However, our work shows that the Oja equation only shows a bifurcation at a critical crosstalk value in very narrow conditions. We suggest that the important analogy lies less in detailed mathematical equivalencies, and more in the fundamental need for accuracy in elementary biological processes. In particular, it's clear that superaccurate polynucleotide copying underlies Darwinian evolution, and similarly superaccurate Hebbian plasticity might be needed for neural learning.

\subsection{Effect of crosstalk on linear learning}

The analysis reported here essentially shows that the well known bifurcation that occurs in linear Hebbian learning as unbiased negative correlations become positive (from segregated to unsegregated states) still occurs in the presence of crosstalk, but at a new, crosstalk-dependent critical negative correlation level. This effect is quite intuitive: crosstalk favors the unsegregated state, and therefore allows the switch to occur at negative correlation, rather than at zero correlation. Of course this situation changes dramatically as soon as any degree of bias is introduced, since now the eigenvalues of $\bf C$ become distinct, and our previous analysis~\cite{radulescu2009hebbian} applies : crosstalk produces a smooth change in the direction of the learned weights (the dominant eigenvector of $\bf EC$). Our present analysis attempts to characterize the relation between these two regimes. In particular, we show that the smooth change can be very rapid when bias is weak.

The change in the normalization produced by crosstalk in the Oja model is largely irrelevant, and indeed one still sees the same behavior with explicit normalization (Section 3). Our analysis also gives insight into the codimension two bifurcation that occurs at the critical quality $q^*$, via an ellipse of half-stable equilibria. The motion towards the  ellipse becomes extremely rapid (Figures 3B and 9), which permits the exchange of stability between the 2 invariant lines (Figure~\ref{invlines}). This rapid motion shows up in simulations (and presumably in biological realizations) as very ``noisy'' weights as the threshold crosstalk value is neared.

Of course a true bifurcation is only seen for unbiased inputs and for negligible learning rates. However, the behavior remains practically indistinguishable from a bifurcation even with slightly biassed inputs and finite learning rates. An example was already discussed in our previous paper (see Figure in ~\cite{radulescu2009hebbian}). A similar situation occurs with models of phase transitions: a true bifurcation of the dynamics only occurs in the ``thermodynamic limit,'' but this is effectively established even for quite small systems~\cite{sollich1994finite}. We have previously called attention to the analogy between Hebbian learning and molecular evolution~\cite{adams2002synaptic,fernando2010neuronal}, with crosstalk playing the role of mutation. In Eigen's evolution model~\cite{eigen1971selforganization}, the transition from the ordered, living, state to the disordered, chemical, state is quite sharp even for polynucleotide lengths $\sim 50$, though a true phase transition (identical to that of the surface of the 2-dimensional Ising model) is only seen with unlimited chains~\cite{saakian2006exact}. Interestingly, the model becomes easiest to analyze in this limit, and the relevant dimensionless control parameter is simply the product of the mutation rate and the (binary) chain length (for binary strings). Although we analyzed here the $n=2$ case, we assume that weights are specified with unlimited bit resolution (i.e., reals). In this case the dimensionless control parameter, equivalent to that in the thermodynamic limit of the Eigen model, is $q$. In the standard Eigen model, the mutation rate is the same at all chain positions. In the next section we discuss the analogous concept for Hebbian learning.

\subsection{The error matrix}

Throughout this paper we assume that the Hebbian adjustment of any weight was equally affected by error, and does not depend either on the strength of that weight or its identity. Such ``isotropicity'' seems a reasonable first assumption, like neglecting bumps on an inclined plane in mechanics. However, it does appear to fly in the face of biological reality. First, stronger synapses are also bigger, and also require higher spine neck conductances and therefore presumably are less well isolated chemically~\cite{koch2005biophysics}. However, such expected ``weight-dependence'' might only be a second-order effect, because in order to ensure that LTP is ``Hebbian,'' the spine neck resistance must always be sufficiently low, even at ``silent,'' AMPAR-less zero-strength synapses, so that the essential back-propagating spike effectively invades the spine head~\cite{magee1997synaptically} during the peak NMDAR opening. Second, crosstalk between individual synapses is a relatively local, not global, phenomenon~\cite{bonhoeffer1989synaptic,harvey2007locally}. However, during learning individual synapses appear and disappear, which will smear details. Furthermore connections are made up of many individual synapses scattered over much of the dendritic tree, which will also smear detail~\cite{radulescu2009hebbian}. Recent work shows that feed-forward cortical connections carrying similar information do not ``cluster'' on dendritic segments, invalidating the argument that local crosstalk could promote useful clustering~\cite{harvey2007locally}.

\subsection{Relevance to ocular dominance and general developmental mechanisms}

A useful though rather fuzzy distinction can be drawn between developmental mechanisms which generate sets of connections (``circuits''), or, perhaps, ``incipient'' or ``potential'' connections~\cite{adams2002new,stepanyants2005neurogeometry} that can be made actual without axo-dendritic rewiring merely by adding postsynaptic spines or presynaptic ``drumsticks''~\cite{anderson2001does,sherman2001exploring}, and ``learning,'' which refines (perhaps in crucial ways) the overall framework established by development. This distinction is related to that between ``Nature'' and ``Nurture,'' or, in the context of Chomskyan linguisitics, ``principles'' and ``parameters.'' The Oja model encapsulates this distinction in minimal form: by definition when the inputs are unbiased there can be no learning, and only two outcomes are possible, which we call segregated or unsegregated. The classic biological example is that in many species early in development a geniculate  axon diffusely innervates a patch of layer IV of cortex (though it does not necessarily contact all the neurons whose dendrites ramify in that patch), but then retracts from stripes within that patch that become selectively innervated by axons corresponding carrying signals from the other eye. Cells within a stripe then becomes largely monocular, although they develop different selectivities for different stimulus features such as orientation. In the Oja model segregation appears in response to unbiased (or, effectively, nearly unbiased) inputs at a critical level of negative correlation, which depends on the degree of crosstalk. In real animals segregation appears before the onset of visual experience, and is thought to be driven by unbiased inputs generated by spontaneous firing.  While one might expect crosstalk to hinder segregation (since it tends to equalize weights), our results show this is not quite correct in the Oja model: it merely shifts the critical degree of (unbiased) correlation required. Various proposals exist for how such inputs can induce segregation even when correlations are positive~\cite{miller1994role,elliott2003analysis} and it's likely that crosstalk will also have the same weak effect here. Indeed, Elliott~\cite{elliott2012} has shown that while crosstalk induces a bifurcation from segregation to unsegregation in a weight-dependent model, the critical value (his equation 3.8) can be shifted to favorable values with suitable correlation values. Thus, the endogenous developmental machinery that creates circuits probably does not require great Hebbian accuracy (and might not require Hebbian machinery at all~\cite{crowley2000early,paik2011retinal}. If the aforementioned postulated layer VI proofreading circuit~\cite{adams2006neurobiological} underlies accuracy, it would not be needed until learning begins, consistent with evidence that the final stages of layer VI circuitry (for example, feedback to relay cells) is late to develop. Indeed, much of the initial pruning that takes place in development might serve to improve the accuracy of proofreading circuitry essential for true learning.

Once detailed, and biased, sensory input occurs, it can drive quantitative adjustments in the already correctly segregated circuits, involving both synapse-strength change and stabilization and un-silencing of new spines (and removal of weak synapses). However, even in the highly simplified Oja model, appropriate adjustment now requires great accuracy, and therefore presumably ``proofreading,'' especially when correlation bias is weak.

\subsection{Normalization and error}

We have assumed in both our papers on the linear Hebb rule that the effect of crosstalk is solely on the Hebbian part of the rule, not the normalizing component. However, if normalization, or some other process that stabilizes Hebbian learning, is biologically necessary, then it is presumably also subject to imperfections such as crosstalk. There are basically two ways to add this other form of crosstalk to stabilized Hebbian learning rules. In the context of a single neuron model, one could simply apply a second, different, crosstalk matrix, say F, to the stabilizing term. However, if ${\bf F}={\bf E}$ (because the geometry underlying such errors is the same in each case) such normalization crosstalk cancels the overall effect. Even if the pattern of crosstalk at each update is described by fluctuating matrices $\mbox{\boldmath ${\cal E}$}$ and $\mbox{\boldmath ${\cal F}$}$, whose averages equal ${\bf E}$ and ${\bf F}$, which do not exactly cancel, one might expect they would on average. Could this be a way to eliminate errors?

The other possible way that errors could affect normalization would be if the underlying normalization mechanism were sufficiently different that the average geometry differed. Although the Oja model allows negative activities, firing rates can only be positive, and it is tempting to suppose that negative signals are carried in special ``off'' channels whose positive weights represent negative ones. In this equivalence, the Hebb part of the rule reflect LTP and the normalization part, LTD. In the cortex, LTP seems to be postsynaptic and LTD presynaptic. If the update leaks presynaptically, within the axon, it will affect a different set of synapses, ones that mostly form onto a different postsynaptic cell. To evaluate how such errors might affect learning, one needs a multi-unit model.

\section{Conclusion}

Generically the inspecific Oja rule does not show bifurcations with variation in the crosstalk parameter. In this paper we analyze an interesting special case which does show a bifurcation: when the input statistics are unbiased. We also describe the behavior in the vicinity of this special case, which is practically indistinguishable from a bifurcation. Essentially in this region ``learning'' changes rather abruptly from being dominated by second-order input statistics (at sufficiently low crosstalk) to being dominated by the internal pattern of crosstalk itself. However, we regard this behavior as being biologically rather uninteresting, since synaptic mechanisms are presumably accurate enough that it never occurs. The one exception would be during development, where near-unbiased statistics might be used by the brain to induce initial selective wiring. Our results suggest that even in this case, high Hebbian accuracy might be required. However, extreme accuracy is probably most essential for nonlinear learning  from higher-order statistics~\cite{cox2009hebbian,elliott2012}.


\bibliographystyle{plain}
\bibliography{arxiv_07_31_12}      


\clearpage

\section*{Appendix 1. A few detailed proofs}

\vspace{3mm}
\begin{em}
\noindent {\bf Lemma.} $Df^{\bf E}_{\bf w} = \gamma \left[ {\bf EC} - 2{\bf w}({\bf Cw})^{T} - ({\bf w}^{T}{\bf Cw}){\bf I} \right]$
\end{em}

\vspace{2mm}
\proof{ Call $g({\bf w})=({\bf w}^{T}{\bf Cw}){\bf w}$ , so $f^{\bf E}({\bf w}) = \gamma [{\bf ECw} - g({\bf w})]$

$$g_{i}({\bf w}) = ({\bf w}^{T}{\bf Cw})w_{i}$$
If $i \not= j$:
$$\frac{\partial g_{i}}{\partial w_{j}}({\bf w}) =
       \frac {\partial}{\partial w_{j}}(\sum_{k,l}C_{kl}w_{k}w_{l})w_{i}
    =2(\sum_{k} C_{kj}w_{k}) w_{i} \, = \, 2[{\bf Cw}]_{j}w_{i}$$
If $i=j$:
$$\frac{\partial g_{i}}{\partial w_{i}}({\bf w}) =
       \frac {\partial}{\partial w_{i}}(\sum_{k,l}C_{kl}w_{k}w_{l})w_{i}
       +\sum_{k,l}C_{kl}w_{k}w_{l}
       =2(\sum_{k}C_{ki}w_{k})w_{i} +$$
     $$ + {\bf w}^{T}{\bf Cw} = 2[{\bf Cw}]_{i}w_{i} +  {\bf w}^{T}{\bf Cw}$$
So: $$Dg_{\bf w} = 2{\bf w}({\bf Cw})^{T} + ({\bf w}^{T}{\bf Cw}){\bf I}$$}

\vspace{3mm}
\begin{em}
\noindent {\bf Proposition 1.2.} Suppose ${\bf EC}$ has a multiplicity one largest eigenvalue. An equilibrium ${\bf w}$ (i.e., by Proposition 1.1, an eigenvector of ${\bf EC}$ with eigenvalue $\lambda_{\bf w}$, normalized so that $\| w \|_{\bf C} = \lambda_{\bf w}$) is a local hyperbolic attractor for the system (1) iff it is an eigenvector corresponding to the maximal eigenvalue of ${\bf EC}$.
\label{attractorthm}
\end{em}

\vspace{2mm}
\proof{Fix an eigenvector ${\bf w}$ of ${\bf EC}$, with ${\bf ECw} = \lambda_{\bf w} {\bf w}$. Then:

\begin{eqnarray*}
Df^{\bf E}_{\bf w}{\bf w} &=& \gamma [{\bf ECw} - 2 {\bf w}({\bf Cw})^{T}{\bf w} - ({\bf w}^{T}{\bf Cw}){\bf w}] = \\
&=& \gamma [- 2 {\bf ww}^{T}{\bf Cw}] = - 2 \gamma \lambda_{\bf w}{\bf w}
\end{eqnarray*}

\noindent Recall that the vector ${\bf w}$ can be completed to a basis ${\cal B}$ of eigenvectors, orthogonal with respect to the dot product $\langle \cdot,\cdot \rangle_{\bf C}$. Let ${\bf v} \in {\cal B}$, ${\bf v} \neq {\bf w}$, be any other arbitrary vector in this basis, so that ${\bf ECv}=\lambda_{\bf v} {\bf v}$, and $\langle {\bf w}, {\bf v} \rangle_{\bf C} = {\bf w}^t{\bf Cv} = 0$. We calculate:

\begin{eqnarray*}
Df^{\bf E}_{\bf w}{\bf v} &=& \gamma [{\bf ECv} - 2 {\bf ww}^{T}{\bf Cv} - \lambda_{\bf w}{\bf v}] = \\
&=& \gamma [(\lambda_{\bf v}-\lambda_{\bf w}){\bf v} -2\langle {\bf w},{\bf v} \rangle _{\bf C}{\bf w}]=- \gamma [\lambda_{\bf w} - \lambda_{\bf v}] {\bf v}
\end{eqnarray*}

\noindent So ${\cal{B}}$ is also a basis of eigenvectors for $Df^{\bf E}_{\bf w}$. The corresponding eigenvalues are $-2 \gamma \lambda_{\bf w}$ (for the eigenvector ${\bf w}$) and $-\gamma[\lambda_{\bf w} - \lambda_{\bf v}]$ (for any other eigenvector ${\bf v} \in {\cal{B}}, \, , \, {\bf v} \not= {\bf w}$). An equivalent condition for ${\bf w}$ to be a hyperbolic attractor for the system (\ref{mothersys}) is that all the eigenvalues of $Df^{\bf E}_{\bf w}$ are $<0$. Since the learning rate $\gamma$ and the eigenvalue $\lambda_{\bf w}$ are both $>0$, this condition is further equivalent to having $- \gamma(\lambda_{\bf w} - \lambda_{\bf v}) \rvert < 0 \, , \text{ for all } {\bf v} \in {\cal{B}} \, , \, {\bf v} \not= {\bf w}$. In conclusion, an equilibrium ${\bf w}$ is a hyperbolic attractor if and only if $\lambda_{\bf w} > \lambda_{\bf v} \, , \, \text{ for all }\, {\bf v} \not= {\bf w}$ \, (i.e. $\lambda_{\bf w}$ is the maximal eigenvalue, or in other words if ${\bf w}$ is in the direction of the principal eigenvector of ${\bf EC}$).
}


\clearpage

\section*{Appendix 2. An extension to higher dimensions}

\vspace{3mm}
\begin{em}
\noindent {\bf Theorem.} Suppose the the modified covariance matrix ${\bf EC}$ has a unique maximal eigenvalue $\lambda_1$. Then the two eigenvectors $\pm {\bf w_{EC}}$ corresponding to $\lambda_1$, normalized such that $\| {\bf w} \|_{\bf C} = \lambda_1$, are the only two attractors of the system. More precisely, the phase space is divided into two basins of attraction, of ${\bf w_{EC}}$ and $- {\bf w_{EC}}$ respectively, separated by the subspace $\langle {\bf w}, {\bf w_{EC}} \rangle = 0$.
\end{em}

\vspace{2mm}
\proof{ We perform the change of variable ${\bf u}=\sqrt{\bf C} {\bf w}$, so that ${\bf u}^t{\bf u} = {\bf w}^t {\bf C} {\bf u}$. Notice that ${\sqrt{\bf C}}$ is also a symmetric matrix, and that ${\bf w} = \sqrt{\bf C}^{-1} {\bf u}$; the system then becomes:

\begin{equation}
\sqrt{\bf C}^{-1} \dot{\bf u} = {\bf EC \sqrt{\bf C}}^{-1} {\bf u} - ({\bf u}^t \sqrt{\bf C}^{-1} {\bf C} \sqrt{\bf C}^{-1} {\bf u}) \sqrt{\bf C}^{-1} {\bf u} = {\bf E \sqrt{\bf C} u} - ({\bf u}^t {\bf u}) \sqrt{\bf C}^{-1} {\bf u} \nonumber
\end{equation}

\noindent or equivalently:

\begin{equation}
\dot{\bf u} = {\bf \sqrt{C} E \sqrt{C} u} - ({\bf u}^t {\bf u}) {\bf u} = {\bf Au} - ({\bf u}^t {\bf u}) {\bf u}
\label{newsys}
\end{equation}

\noindent where, of course, we defined ${\bf A = \sqrt{C} E \sqrt{C}}$. Clearly, ${\bf A}$ a symmetric matrix, having the same eigenvalues as ${\bf EC}$. More precisely, ${\bf w}$ is an eigenvector of ${\bf EC}$ with eigenvalue $\mu$ iff ${\bf \sqrt{C} v}$ is an eigenvector of ${\bf A}$ with eigenvalue $\mu$. Moreover: any two distinct eigenvectors ${\bf v} \neq {\bf w}$ of ${\bf EC}$ are known to be orthogonal, hence any two distinct eigenvectors of ${\bf A}$ are orthogonal in the regular Euclidean dot product: $({\bf \sqrt{C}v})^t({\bf \sqrt{C}w}) = {\bf v}^t \sqrt{\bf C} \sqrt{\bf C} {\bf w} = {\bf v}^t {\bf Cw} = 0$.\\

\noindent Consider then ${\bf v}$ to be the principal component of ${\bf A}$ (i.e., the eigenvector corresponding to its maximal eigenvalue), and let ${\bf u = u}(t)$ be a trajectory of the system (\ref{newsys}). We want to observe the evolution in time of the angle between the variable vector ${\bf u}$ and the fixed vector ${\bf v}$.

\begin{equation}
\cos {\theta} = \frac{\langle {\bf v}, {\bf u} \rangle}{\| {\bf v} \| \cdot |\ {\bf u} \|}
\nonumber
\end{equation}

\noindent We differentiate and obtain:

\begin{equation}
- \| {\bf v} \| \sin(\theta) \dot{\theta} = \frac{1}{\| {\bf u} \|^2} \left[ \langle {\bf v}, \dot{\bf u} \rangle \cdot \| {\bf u} \| - \langle {\bf v}, {\bf u} \rangle  \frac{\langle {\bf u}, \dot{\bf u} \rangle}{\| {\bf u} \|} \right] = \frac{({\bf v}^t \dot{\bf u}) \| {\bf u}\|^2 - ({\bf v}^t{\bf u}) ({\bf u}^t {\bf u})}{\| {\bf u} \|^3}
\end{equation}

\noindent The numerator of this expression

\begin{eqnarray*}
h({\bf u}) &=& ({\bf v}^t \dot{\bf u}) ({\bf u}^t {\bf u}) - ({\bf v}^t {\bf u}) ({\bf u}^t \dot{\bf u})
 = ({\bf u}^t {\bf u}) \Bigl( {\bf v}^t [ {\bf Au} - ({\bf u}^t {\bf u}) {\bf u}] \Bigr) + ({\bf u}^t {\bf v}) - ({\bf v}^t {\bf u}) \Bigl( {\bf u}^t - [ {\bf Au} - ({\bf u}^t {\bf u}) {\bf u}] \Bigr)\\
 &=& ({\bf u}^t {\bf u})({\bf v}^t {\bf Au}) - ({\bf v}^t {\bf u})({\bf u}^t {\bf Au})
\end{eqnarray*}

\noindent We are interested in the sign of $h({\bf u})$; to make our computations simpler, we can diagonalize ${\bf A}$ in a basis of orthogonal eigenvectors  ${\bf A} = {\bf P}^t{\bf DP}$, where ${\bf D}$ is the diagonal matrix of eigenvalues and ${\bf P}$ is an orthogonal matrix whose columns are the eigenvectors. Then:

\begin{eqnarray*}
h({\bf u}) &=&
[({\bf Pu}^t)({\bf Pu})][({\bf Pv}^t){\bf D}({\bf Pu})] - [({\bf Pv}^t)({\bf Pu})][({\bf Pu}^t){\bf D}({\bf Pu})]\\
&=& ({\bf z}^t {\bf z}) ({\bf y}^t {\bf D}{\bf z}) - ({\bf y}^t {\bf z}) ({\bf z}^t {\bf D}{\bf z})
\end{eqnarray*}

\noindent where ${\bf y} = {\bf Pv}$ and ${\bf z} = {\bf Pu}$, so that ${\bf Dy} = {\bf DPv} = \lambda_1 {\bf y}$ (where $\lambda_1 > \lambda_2 \geq \hdots \geq \lambda_n$ is the largest eigenvalue of ${\bf EC}$, assumed to have multiplicity one. Hence:

\begin{eqnarray*}
h({\bf u}) &=& ({\bf z}^t {\bf z}) ({\bf y}^t {\bf D}{\bf z}) - ({\bf y}^t {\bf z}) ({\bf z}^t {\bf D}{\bf z}) =  ({\bf z}^t {\bf z}) \lambda_1 ({\bf y}^t {\bf z}) - ({\bf y}^t {\bf z}) ({\bf z}^t {\bf D}{\bf z}) = ({\bf y}^t {\bf z}) [\lambda_1 ({\bf y}^t {\bf z}) - {\bf z}^t {\bf D}{\bf z}]\\
&=& ({\bf y}^t {\bf z}) \left[ \lambda_1 \sum{z_j^2} - \sum{\lambda_j z_j^2} \right]
= ({\bf y}^t {\bf z}) \left[ \sum{(\lambda_1 - \lambda_j) z_j^2} \right]
\end{eqnarray*}

\noindent Hence, if ${\bf y}^t {\bf z} >0$, then $h({\bf u}) > 0$. In other words: if ${\bf v}^t {\bf u} > 0$ then $- \| {\bf v} \| \sin(\theta) \dot{\theta} > 0$, hence that $\dot{\theta} < 0$. For our original system, this means that any trajectory starting at a ${\bf w}$ with $\langle {\bf w}, {\bf w_{EC}} \rangle > 0$ converges in time towards the principal eigenvector ${\bf w_{EC}}$ of the matrix ${\bf EC}$.

}


\clearpage

\section*{Appendix 3. Sensitivity analysis}

This is a technical section, in which we calculate how the invariant directions $z_{1,2}$ change when varying $q$.\\

\noindent {\bf Remark.} In order to simplify further computations, we rewrite:

\begin{eqnarray*}
z_{1,2} &=& \frac{-q \delta \pm \sqrt{\Delta}}{2 \beta} = \frac{-q \delta \pm \sqrt{q^2 \delta^2 + 4\beta[\beta+(1-q)\delta]}}{2 \beta}\\
&=& -\frac{1}{2} \left( \frac{q \delta}{\beta} \right) \pm \frac{1}{2} \text{sign}(\beta) \sqrt{\left( \frac{q \delta}{\beta}\right)^2 + \frac{4[\beta+(1-q)\delta]}{\beta}}
\end{eqnarray*}

\noindent Call $\displaystyle{\gamma = \frac{q \delta}{\beta} = \frac{q \delta}{cq+(1-q)v}}$. Then $\displaystyle{\frac{(1-q)\delta}{\beta}=\frac{\delta-c \gamma}{v}}$, and hence:

$$z_{1,2} = -\frac{1}{2} \gamma \pm \frac{1}{2} \text{sign}(\beta) \sqrt{\eta}$$

\noindent where

$$\eta=\frac{\Delta}{\beta^2}=\gamma^2+4 \left[ 1+\frac{\delta-c \gamma}{v} \right]$$

Then we can use the chain rule to express $\displaystyle{\frac{dz_{1,2}}{dq} = \frac{dz_{1,2}}{d \gamma} \cdot \frac{d \gamma}{dq}}$.

\vspace{3mm}
\begin{em}
\noindent {\bf Lemma.} The derivative $\displaystyle{\frac{d \gamma}{dq} = \frac{\delta v}{\beta^2}}$. Also, for $q\in (1/2,q^*) \cup (q^*,1]$ (i.e., where $\beta \neq 0$), we have:
$$\frac{dz_{1,2}}{d \gamma}= \frac{1}{2} \left[ \frac{\displaystyle{\pm \left( q\delta - \frac{2c\beta}{v} \right)}}{\sqrt{\Delta}} - 1 \right]<0$$
\end{em}

\vspace{2mm}
\proof{ $\displaystyle{\frac{d \gamma}{dq} = \frac{d}{dq} \left(  \frac{q \delta}{\beta} \right) = \frac{\delta[\beta - q \dot{\beta}]}{\beta^2} = \frac{\delta[\beta - q(c-v)]}{\beta^2} = \frac{\delta v}{\beta^2}}$\\

\noindent For $q\in (1/2,q^*) \cup (q^*,1]$, we also have directly that:

\begin{eqnarray}
\frac{dz_{1,2}}{d \gamma} &=& = -\frac{1}{2} \pm \frac{1}{2} \frac{\displaystyle{\text{sign}(\beta) \frac{d\eta}{d\gamma}}}{\sqrt{\eta}}
= -\frac{1}{2} \pm \frac{1}{2} \text{sign}(\beta) \frac{\displaystyle{\gamma - \frac{2c}{v}}}{\sqrt{\displaystyle{\gamma^2+4 \left[ 1+\frac{\delta-c \gamma}{v} \right]}}}\\
&=& \frac{1}{2} \left[ \pm \frac{\displaystyle{\beta \left( \gamma-\frac{2c\beta}{v} \right) }}{\sqrt{\Delta}/\lvert \beta \rvert} - 1 \right]
= \frac{1}{2} \left[ \frac{\displaystyle{\pm \left( q\delta - \frac{2c\beta}{v} \right) }}{\sqrt{\Delta}} - 1 \right]
\end{eqnarray}

\noindent Since $\displaystyle{\left( \gamma^2+4 \left[ 1+\frac{\delta-c \gamma}{v} \right] \right) -\left( \gamma - \frac{2c}{v} \right)^2 = \frac{4[v(v+\delta)-c^2]}{v^2} > 0}$, it follows that :

$$\sqrt{\gamma^2+4 \left[ 1+\frac{\delta-c \gamma}{v} \right] } > \lvert \gamma -\frac{2c}{v} \rvert \geq \pm \left( \gamma -\frac{2c}{v} \right)$$

\noindent and hence

$$\frac{\displaystyle{\left\lvert \text{sign}\beta \left( \gamma - \frac{2c}{v} \right) \right\rvert}}{\sqrt{\displaystyle{\gamma^2+4 \left[ 1+\frac{\delta-c \gamma}{v} \right]}}} < 1$$

\noindent It immediately follows from (4) that $\displaystyle{\frac{dz_{1,2}}{d \gamma}<0}$.
}

\vspace{3mm}
\begin{em}
\noindent {\bf Corollary.} The slope of the invariant lines changes with respect to $q$ according to:
$$\frac{dz_{1,2}}{dq}= \frac{\delta v}{2 \beta^2} \left[ \frac{\pm \left( \displaystyle{q\delta - \frac{2c\beta}{v}} \right)}{\sqrt{\Delta}} - 1 \right]$$
\noindent hence $\displaystyle{\emph{sign} \left( \frac{dz_{1,2}}{dq} \right) = -\emph{sign}(\delta)}$ for $q\in (1/2,q^*) \cup (q^*,1]$.
\end{em}

\vspace{2mm}
\proof{The conclusion follows directly from the chain rule that $\displaystyle{\frac{dz_{1,2}}{dq} = \frac{\delta v}{\beta^2} \cdot \frac{dz_{1,2}}{d \gamma}}$.}

At this stage, we can distinguish two cases: $\delta<0$ and $\delta>0$. We analyze in detail the case $\delta>0$. The other is very similar (although not symmetric about $\delta=0$), and we will only state the results, and show some graphic illustrations.

\vspace{3mm}
\begin{em}
\noindent {\bf Proposition 2.6.} If $\delta>0$, then $\displaystyle{\frac{d z_{1,2}}{dq}}<0$; hence both $z_{1,2}$ are decreasing as $q \in (1/2, q^*) \cup (q^*,1]$. Furthermore, the monotonicity, asymptotes and end behavior of the functions $z_{1,2}(q)$ are sketched in the following table:

\begin{center}
\begin{tabular}{l|lcccc}
$q$ & $1/2$ & & $q^*$ & & $1$\\
\hline
$z_1$ & $1$ & $\quad \searrow \quad$ & $\displaystyle{\frac{1-q^*}{q^*}}$ & $\quad \searrow \quad$ & $\displaystyle{\frac{\delta-\sqrt{4c^2+\delta^2}}{-2c}}$\\
\hline
$z_2$ & $\displaystyle{ -1-\frac{\delta}{2(v+c)} }$ & $\searrow$ & $_{-\infty}\vert^{\infty}$ & $\searrow$ & $\displaystyle{\frac{\delta+\sqrt{4c^2+\delta^2}}{-2c}}$\\
\end{tabular}
\end{center}
\end{em}

\noindent {\bf Remark 1.} In the system's phase plane, this corresponds to a continuous clockwise rotation of the two invariant lines (the vertical asymptote at $q^*$ corresponds to the $z_2$ line going through a the vertical position). A phase-plane sketch of this process is shown in Figure 2 in the main text, and the graphs of the actual functions $z_{1,2}(q)$ and of their derivatives $d z_{1,2}/dq$, for some fixed values of the parameters $v,c,\delta > 0$, are shown in the Figure below.\\

\begin{figure}[h!]
\includegraphics[scale=0.23]{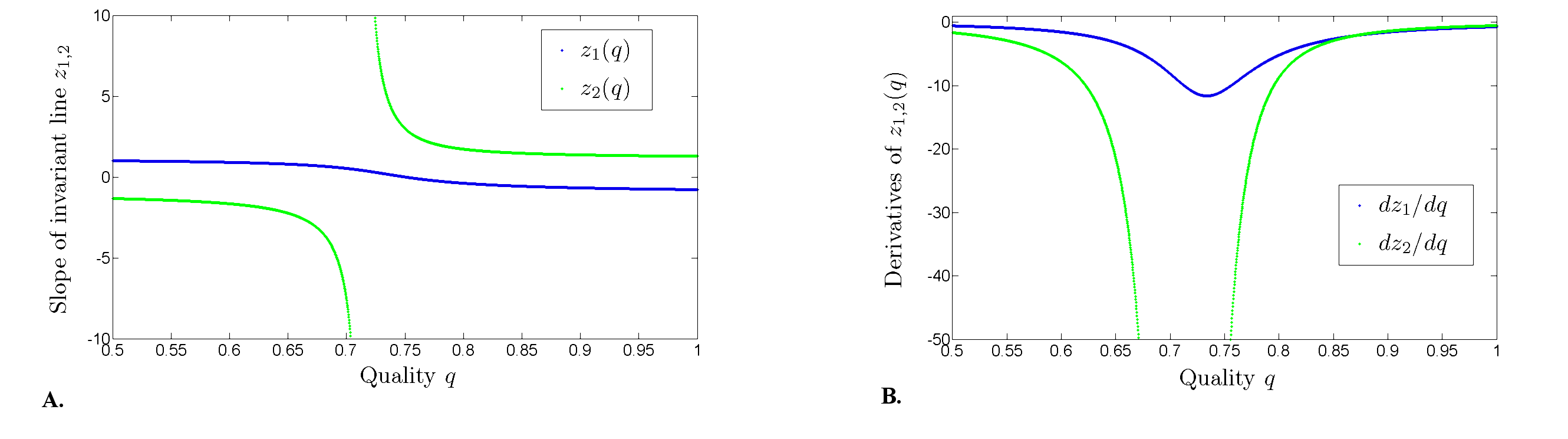}
\noindent {\bf Figure: Slopes of invariant lines (A) and their change as $q$ is varied (B).} In both panels, the other parameters values were fixed to $v=1$, $c=-0.4$ and $\delta=0.2$. Notice that, in accordance with Proposition 2.6, $z_1$ and its derivative $d z_1/dq$ are continuous (blue curves) on $[1/2,1]$, while $z_2$ and its derivative $d z_2/dq$ (green curves) have vertical asymptotes at $\displaystyle{q=q^*=\frac{v}{v-c} \sim 0.71}$.
\label{slopes}
\end{figure}

\noindent {\bf Remark 2.} Clearly from the table, the angular position of the two equilibria at $q=q^*$ does not depend on the bias $\delta$. It can be easily shown that the norm of these points is also independent on $\delta$. For example, the norm of the stable equilibrium is:

\begin{eqnarray}
\lVert {\bf w} \rVert^2 &=& \frac{\mu_1(z_1^2+1)}{vz_1^2+2cz_1+(v+\delta)}=\frac{(1-q^*)c+q^*(v+\delta)}{v(1-q^*)^2+2cq^*(1-q^*)+(v+\delta)q^{*^2}} \cdot q^{*^2}(z_1^2+1)\\
&=& \frac{(1-q^*)c+q^*(v+\delta)}{q^*[(1-q^*)c+q^*(v+\delta)]} \cdot q^{*^2}(z_1^2+1) = \frac{1-2q^*+2q^{*^2}}{q^*}
\end{eqnarray}

\noindent Hence the position of the two equilibria at critical quality does is the same for all bias values $\delta>0$.

\vspace{2mm}
\proof{The monotonicity follows from the Corollary. The limit values follow from direct computation. For example:\\

$\displaystyle{\lim_{q \to q^*_+}{z_2} = \lim_{q \to q^*_+}{\frac{-q \delta - \sqrt{\Delta}}{2 \beta}} = \lim_{q \to q^*_+}{\frac{-q^* \delta}{0^-}} = +\infty}$\\

$\displaystyle{\lim_{q \to q^*_-}{z_2} = \lim_{q \to q^*_-}{\frac{-q^* \delta}{0^+}} = -\infty}$\\

$\displaystyle{\lim_{q \to q^*}{z_1} = \lim_{q \to q^*}{\frac{-q \delta + \sqrt{\Delta}}{2 \beta} \cdot \frac{-q \delta - \sqrt{\Delta}}{-q \delta - \sqrt{\Delta}}} = \lim_{q \to q^*}{\frac{\beta+(1-q)\delta}{q \delta + \sqrt{\Delta}}} = \frac{1-q^*}{q^*}}$
}

\noindent {\bf Remark 3.} If $\delta<0$, then $\displaystyle{\frac{d z_{1,2}}{dq}>0}$ and hence both $z_1$ and $z_2$ are increasing as $q \in (1/2, q^*) \cup (q^*,1]$. In the system's phase plane, this corresponds to a continuous counter-clockwise rotation of the two invariant lines.

\vspace{3mm}
\begin{em}
\noindent {\bf Proposition 2.7.} For $\delta>0$, the angle $\theta_{1,2} \in [-\pi/2,\pi/2]$ between each invariant line and the $w_1$ abscissa is decreasing with respect to the parameter $q$. Moreover, the angular rate of change is finite, at all $q\in (1/2,1]$.

\begin{center}
\begin{tabular}{l|lcccc}
$q$ & $\displaystyle{\frac{1}{2}}$ & & $q^*$ & & $1$\\
\hline
$d\theta_1/dq$ & & $(-)$ & $\displaystyle{-\frac{\det(C)}{v \delta (1-2*q+2q^2)}}$  & $(-)$ & \\
\hline
$d\theta_2/dq$ &  & $(-)$ & $\displaystyle{-\frac{v}{q^2 \delta}}$ & $(-)$ & \\
\end{tabular}
\end{center}
\end{em}

\vspace{2mm}
\proof{The relation between the slope $z$ and the actual angle $\theta$ is given by: $z = \tan{\theta}$ (we will avoid indices wherever there is no danger of confusion). Hence, for $q \in (1/2, q^*) \cup (q^*,1]$, we have:

$$\cos^2(\theta) \cdot \frac{d\theta}{d\gamma} \; \Longrightarrow \;  \frac{d\theta}{d\gamma}=\frac{dz}{d\gamma} \cdot \frac{dz}{d\gamma} \cdot \frac{1}{z^2+1}$$.

\noindent So

\begin{eqnarray}
\frac{d\theta}{dq}=\frac{d\gamma}{dq} \cdot \frac{d\theta}{d\gamma}=\frac{\delta v}{2 \beta^2} \cdot \frac{dz}{d\gamma} \cdot \frac{1}{z^2+1}
\end{eqnarray}

\noindent hence $\displaystyle{\text{sign} \left( \frac{d\theta}{dq} \right) = -\text{sign} (\delta)}$, for all $q\in (1/2,q^*) \cup (q^*,1]$.

We yet have to check that the rate of change $\displaystyle{\frac{d\gamma}{dq}}$ remains finite (i.e., does not blow up to $-\infty$) as $q \to q^*$. Elaborating on (6) we have, for $q\in (1/2,q^*) \cup (q^*,1]$:

\begin{eqnarray}
\lim_{q \to q^*}{\frac{d \theta_2}{dq}} &=& \lim_{q \to q^*}{\frac{\delta v}{2 \beta^2} \cdot \frac{- \left( q \delta - \frac{2c \beta}{v} \right) - \sqrt{\Delta}}{\sqrt{\Delta}} \cdot \frac{4 \beta^2}{\Delta + 4 \beta^2 - 2q \delta \sqrt{\Delta}}} \nonumber \\
&=& \frac{2 \delta v}{q^* \delta} \cdot \frac{-q^* \delta - q^* \delta}{2q^{*^2} \delta^2 + 2q^{*^2} \delta^{*^2}} = -\frac{v}{q^{*^2} \delta}
\end{eqnarray}

\noindent We also notice that\\

\begin{equation}
\lim_{q \to q^*}{z_1} = \frac{1-q^*}{q^*} \; \Longrightarrow \; \lim_{q \to q^*}{(z_1^2+1)} = \frac{1-2q^*+2q^{*^2}}{q^{*^2}}
\end{equation}

\noindent and that

\begin{eqnarray}
\lim_{q \to q^*}{\frac{dz_1}{dq}} &=& \lim_{q \to q^*}{\frac{\delta v}{2 \beta^2} \cdot \frac{\left( \displaystyle{q \delta - \frac{2c \beta}{v}} \right) - \sqrt{\Delta}}{\sqrt{\Delta}}} = \lim_{q \to q^*}{\frac{\delta v}{2 \beta^2} \cdot \frac{\left( \displaystyle{q \delta - \frac{2c \beta}{v}} \right) - \sqrt{\Delta}}{\sqrt{\Delta}} \cdot \frac{\left( \displaystyle{q \delta - \frac{2c \beta}{v}} \right) + \sqrt{\Delta}}{\left( \displaystyle{q \delta - \frac{2c \beta}{v}} \right) + \sqrt{\Delta}}} \nonumber \\
&=& \lim_{q \to q^*}{\frac{\delta v}{2 \beta^2 \sqrt{\Delta}} \cdot \frac{\left( \displaystyle{q \delta - \frac{2c \beta}{v}} \right)^2 - \Delta}{\left( \displaystyle{q \delta - \frac{2c \beta}{v}} \right) + \sqrt{\Delta}}} = \lim_{q \to q^*}{\frac{\delta v}{2 \beta^2 \sqrt{\Delta}} \cdot \frac{\displaystyle{\frac{-4 \beta^2}{v^2}} (c^2-v^2-v \delta)}{\left( \displaystyle{q \delta - \frac{2c \beta}{v}} \right) + \sqrt{\Delta}}} \nonumber \\
&=& \lim_{q \to q^*}{\frac{\delta v}{2 \beta^2 \sqrt{\Delta}} \cdot \frac{-4 \beta^2 \det({\bf C})}{v^2} \cdot \frac{1}{\displaystyle{\left( q \delta - \frac{2c \beta}{v} \right)} + \sqrt{\Delta}}} \nonumber \\
&=& \frac{\delta v}{2 \beta^2 q^* \delta} \cdot \frac{-4 \beta^2 \det({\bf C})}{v^2} \cdot \frac{1}{2q^* \delta} = \frac{-\det({\bf C})}{v \delta q^{*^2}}
\end{eqnarray}

\noindent Combining (9) and (10), we have:

\begin{equation}
\lim_{q \to q^*}{\frac{d \theta_2}{dq}} = \frac{dz_1}{dq} \frac{1}{z_1^2+1} = \frac{-\det({\bf C})}{v \delta q^{*^2}} \cdot \frac{q^{*^2}}{1-2q^*+2q^{*^2}} = \frac{-\det({\bf C})}{v \delta (1-2q^*+2q^{*^2})}
\end{equation}

}


\clearpage

\section*{Appendix 4. Description of the ellipse attractor}

For unbiased inputs $\delta=0$ and critical quality $q=q^*$, ${\bf EC}$ has a double eigenvalue $\mu=v+c=(2q^*-1)(v-c)$. The eigenspace of ${\bf EC}$ is $\mathbb{R}^2$, hence each direction (i.e. slope $z = \tan\theta \in [-\infty,+\infty]$) produces two equilibria, normalized as follows:

\begin{eqnarray}
\lVert {\bf w} \rVert^2 &=& \frac{\mu(z^2+1)}{vz^2+2cz+v} = \frac{(v+c) \left[ \tan^2\theta+1 \right]}{v\tan^2\theta+2c\tan\theta+v} = \frac{v+c}{v\sin^2\theta+2c\sin\theta\cos\theta+v\cos^2 \theta} \nonumber \\
&=& \frac{v+c}{v+c\sin(2\theta)}
\end{eqnarray}

\noindent We show that this is the polar equation of an ellipse with foci along the first diagonal $\theta=\pi/4$. Indeed, under a clockwise rotation by $-\pi/4$, equation (13) becomes:

\begin{eqnarray}
\rho^2 &=& \frac{v+c}{v+c\sin \left(2 \left[\theta-\frac{\pi}{4} \right] \right)} = \frac{v+c}{v+c\cos(2\theta)} = \frac{v+c}{v \left[ \cos^2\theta+\sin^2\theta \right] + c \left[ \cos^2\theta-\sin^2\theta \right]} \nonumber \\
&=& \frac{v+c}{(v+c)\cos^2\theta + (v-c)\sin^2\theta} = \frac{v+c}{\sqrt{v^2-c^2}} \cdot \frac{\sqrt{v^2-c^2}}{(v+c)\cos^2\theta+(v-c)\sin^2\theta}
\end{eqnarray}

\noindent In polar coordinates, this is the equation of an ellipse

$$\rho^2 = \frac{a^2b^2}{a^2\cos^2\theta+b^2\sin^2\theta}$$

\noindent with radial coordinate $\displaystyle{\rho = \lVert {\bf w} \rVert \frac{\sqrt{v^2-c^2}}{v+c}}$ and angular coordinate $\theta$, semi-major radius $a=\sqrt{v+c}$ and semi-minor radius $b=\sqrt{v-c}$.

\end{document}